\documentclass[sn-basic, Numbered]{sn-jnl}
\usepackage{mathptmx}
\usepackage{amsmath,amssymb,amsfonts}%
\usepackage{amsthm}%
\usepackage{eucal}
\usepackage{subcaption}

\DeclareMathOperator{\E}{E}
\DeclareMathOperator{\Var}{Var}

\newcommand{\RR}{\theta}
\newcommand{\LRR}{\Theta}
\newcommand{\OR}{\psi}
\newcommand{\LOR}{\Psi}
\newcommand{\parg}{\zeta} 
\newcommand{\vahparg}{\hat \zeta} 
\newcommand{\suc}{r} 
\newcommand{\effic}{\eta}
\newcommand{\tarvar}{A} 
\newcommand{\roprpr}{\phi}
\newcommand{\roprprsup}{\roprpr_0}
\newcommand{\harm}[1]{H_{#1}}
\newcommand{\CR}{Cram{\'e}r--Rao}
\newcommand{\vanii}{T} \newcommand{\vanoi}{U} \newcommand{\vanoo}{V} \newcommand{\nii}{t} \newcommand{\noi}{u} \newcommand{\noo}{v} 
\newcommand{\vai}{X} \newcommand{\vao}{Y}
\newcommand{\sei}{\mathcal X} \newcommand{\seo}{\mathcal Y}
\newcommand{\sef}{\sigma} 
\newcommand{\vef}{\tau} 
\newcommand{\adjsuc}{\alpha} 
\newcommand{\cerr}{\mu}
\newcommand{\prmax}{\rho}
\newcommand{\prmaxsup}{\prmax_0}
\newcommand{\efficm}{\omega} 
\newcommand{\sefm}{\omega_{\sef}} 

\newcommand{\tamdosfig}{.499\textwidth}
\newcommand{\legendscale}{.58}

\makeatletter
\newcommand\cond{
 \@ifstar
  {\mathrel{}\middle|\mathrel{}}
  {\mid}%
}
\makeatother

\theoremstyle{plain}
\newtheorem{proposition}{Proposition}
\newtheorem{theorem}{Theorem}

\theoremstyle{definition}
\newtheorem{algorithm}{Algorithm}

\graphicspath{{figures/}}

\begin{document}

\title{
Efficient estimation of relative risk, odds ratio\\ and their logarithms for rare events
}


\author[*]{\fnm{Luis} \sur{Mendo}}\email{luis.mendo@upm.es}

\affil[*]{\orgdiv{Information Processing and Telecommunications Center}, \orgname{Universidad Polit\'ecnica de Madrid}, \orgaddress{\street{Avenida Complutense, 30}, \city{Madrid}, \postcode{28040}, \country{Spain}}}


\abstract{
This paper proposes sequential estimators for the relative risk, odds ratio, log relative risk or log odds ratio of a dichotomous attribute in two populations. The estimators take the same number of observations from each population, and ensure that the relative mean-square error for the relative risk or odds ratio, or the mean-square error for their logarithmic versions, is less than a given target, regardless of the value of the estimated parameter. The estimation efficiency, computed with respect to the \CR{} bound, is high when the considered attribute is rare or moderately rare.
}



\maketitle

\section{Introduction}
\label{part: intro}

Consider two populations with probabilities $p_1$ and $p_2$ of occurrence of a certain dichotomous attribute. The \emph{relative risk} (RR) or \emph{risk ratio}, $p_1/p_2$, and the \emph{odds ratio} (OR), $p_1(1-p_2)/(p_2(1-p_1))$, are commonly used measures of association between the prevalence of the attribute in the two populations. They find widespread use in many branches of medical and social sciences, such as epidemiology and psychology \cite{Agresti02}. Also used often are their logarithmic versions: \emph{log relative risk} (LRR), $\log(p_1/p_2)$, and \emph{log odds ratio} (LOR), $\log(p_1(1-p_2)/(p_2(1-p_1)))$ \cite{Armitage02}.

When estimating any of these parameters, it is desirable to \emph{guarantee} a given accuracy of the estimation. Accuracy is often defined in terms of \emph{mean-square-error} (MSE) or \emph{root-mean-square-error} (RMSE). As argued in previous works by the author \cite{Mendo25b, Mendo25c}, for RR and OR it is meaningful to aim for a certain level of \emph{relative} accuracy, such as RMSE divided by the actual value of the parameter (or MSE divided by the square of the parameter); whereas for LRR and LOR the RMSE (or MSE) is an appropriate measure of accuracy, because the logarithm inherently provides a normalization by transforming ratios into differences. In the following, the target accuracy will be assumed to be specified in terms of relative MSE for RR and OR, or MSE for LRR and LOR; and in both cases it will be denoted as $\tarvar$.

This work focuses on estimating the above parameters from sequential observations of the two populations. Samples are considered to be taken \emph{in pairs} consisting of one sample from each population.
This is a particular case of group sequential sampling \cite{Pocock77}. The samples are modelled as Bernoulli random variables, and are assumed to be statistically independent. Specifically, observations from population $i=1,2$ are represented as a sequence $\sei_i$ of independent Bernoulli random variables with parameter $p_i \in (0,1)$.

The estimators presented in the previous work \cite{Mendo25c} can achieve an exact ratio of sample sizes using group sampling, and can thus be particularized to the setting studied in this paper, namely by considering groups consisting of one sample from each population. These estimators guarantee that the relative MSE for RR and OR, or the MSE for LRR and LOR, is less than a target value. They use a form of sequential sampling, based on \emph{inverse binomial sampling} (IBS) \citetext{\citealp{Haldane45}; \citealp[chapter~2]{Lehmann98}}.
This approach extends the method introduced in \cite{Mendo25b} to estimate the \emph{odds} $p/(1-p)$, or the \emph{log odds} $\log(p/(1-p))$, for a single population with parameter $p$.

The approach used in this paper is based on a different way of extending the methodology in \cite{Mendo25b} to two populations. Namely, by observing samples from the two populations, independent Bernoulli random variables with a certain parameter $p$ can be generated such that the odds $p/(1-p)$ equal the RR $p_1/p_2$ or the OR $p_1(1-p_2)/(p_2(1-p_1))$; and then the odds or log odds estimators from \cite{Mendo25b} can be applied to these variables. The resulting estimators are unbiased and guarantee a given accuracy in terms of relative MSE for RR and OR, or MSE for LRR and LOR, for any $p_1, p_2 \in (0,1)$. Moreover, they turn out to have very high efficiency, in particular better than that in \cite{Mendo25c}, for $p_1, p_2$ small.

The main limitation of the method presented in this work is that it considers that samples are taken in pairs, one sample from each of the two populations. This is, however, a very common sampling scenario. Indeed, sampling in pairs has been studied in a large number of references, covering a variety of settings under different assumptions; see for example the works by \citet{Siegmund82, Cho19, Cho20, Kokaew23}. The contribution of this paper is that the proposed estimation method guarantees that a target accuracy is achieved for any $p_1, p_2 \in(0,1)$, while ensuring very good efficiency when these probabilities are small.

The following notation and elementary identities will be used throughout the paper. The two possible outcomes of a Bernoulli random variable, $1$ and $0$, will be respectively called ``success'' and ``failure'', as usual. Geometric random variables are defined starting at value $1$. Thus, a geometric random variable $L$ 
with parameter $p$ has probability function $\Pr[L=k] = (1-p)^{k-1}p$, $k \geq 1$, and $\E[L] = 1/p$. A negative binomial random variable $N$ 
with parameters $\suc$ and $p$ is defined as the number of independent Bernoulli trials with parameter $p$ that are required to obtain exactly $\suc$ successes; and then
\begin{align}
\label{eq: E N neg bin}
\E[N] &= \frac{\suc} p, \\
\label{eq: E Var neg bin}
\Var[N] & = \frac{\suc(1-p)} {p^2}.
\end{align}
The binomial and multinomial coefficients are denoted as
\begin{align}
\binom{n}{k} &= \frac{n!}{k! (n-k)!}, \quad n \geq k \geq 0,\\
\binom{n}{k_1, \ldots, k_m} &= \frac{n!}{k_1! \cdots k_m!}, \quad m \geq 2, \ n \geq k_j \geq 0, \ j=1,\ldots, m.
\end{align}
The harmonic numbers are denoted as $\harm{k} = 1 + 1/2 + \cdots + 1/k$, $k \in \mathbb N$. The following identity is obtained differentiating the geometric series with ratio $s$, twice:
\begin{equation}
\label{eq: geom series diff diff}
\sum_{k=2}^\infty k(k-1) s^{k-2} = \frac 2 {(1-s)^3}.
\end{equation}

The rest of the paper is structured as follows. Section~\ref{part: estim proc} describes the estimation procedure and discusses basic properties of the estimators. Section~\ref{part: E max, gen} characterizes the average number of input pairs used by the estimators. Section~\ref{part: effic} analyses the estimation efficiency and provides lower bounds. Section~\ref{part: concl} presents the conclusions of this work. Appendix~\ref{part: proofs} contains proofs to all results.

\section{Estimation procedure}
\label{part: estim proc}

Estimating the RR $p_1/p_2$ or the LRR $\log(p_1/p_2)$ is equivalent to estimating the odds $p/(1-p)$ or the log odds $\log(p/(1-p))$ if $p$ is suitably chosen, as mentioned in Section~\ref{part: intro}; namely if
\begin{equation}
\label{eq: p RR LRR}
p = \frac{p_1}{p_1+p_2}. 
\end{equation}
A simple method to generate a sample $Y$ with Bernoulli parameter $p$, as defined in Eq.~\eqref{eq: p RR LRR}, using samples from sequences $\sei_1$ and $\sei_2$ is given next. This algorithm (as well as that which will be introduced later for OR and LOR) is an instance of a multiparameter Bernoulli factory \cite{PaesLeme23}.

\begin{algorithm}[Probability transformation for RR and LRR]
\label{algo: RR LRR, inner}
\ \\
\textbf{Inputs}: As many samples from sequences $\sei_1$, $\sei_2$ as needed.
\begin{enumerate}
\item
\label{step: inner RR LRR: go back}
Choose $i=1$ or $2$ equally likely and independently from other variables.
\item
\label{step: inner RR LRR: input needed}
Take a sample $\vai_i$ from sequence $\sei_i$.
\item
If $\vai_i=0$, go to step~\ref{step: inner RR LRR: go back}. Else, set $\vao=1$ if $i=1$ or $\vao=0$ if $i=2$.
\end{enumerate}
\textbf{Output}: $\vao$.
\end{algorithm}

Let $\vanii_1$ and $\vanii_2$ denote the numbers of samples from $\sei_1$ and $\sei_2$ used by one execution of Algorithm~\ref{algo: RR LRR, inner}. The variables $\vanii_1$ and $\vanii_2$ are statistically dependent: large values of one tend to occur when the other also takes large values, as is clear from the definition of the algorithm. The next proposition establishes that Algorithm~\ref{algo: RR LRR, inner} indeed produces the desired output, and gives several identities for $\vanii_1$ and $\vanii_2$ that will be useful to derive subsequent results.

\begin{proposition}
\label{prop: inner RR LRR}
Algorithm~\ref{algo: RR LRR, inner} terminates with probability $1$. Its output has $\Pr[\vao=1] = p$, $\Pr[\vao=0] = 1-p$, with $p$ given by Eq.~\eqref{eq: p RR LRR}. In addition,
\begin{align}
\label{eq: E vanii 1 cond vao 1}
\E[\vanii_1 \cond \vao=1] & = \frac{1+p_2}{p_1+p_2}, \\
\label{eq: E vanii 1 cond vao 0}
\E[\vanii_1 \cond \vao=0] & = \frac{1-p_1}{p_1+p_2}, \\
\label{eq: E vanii 2 cond vao 1}
\E[\vanii_2 \cond \vao=1] & = \frac{1-p_2}{p_1+p_2}, \\
\label{eq: E vanii 2 cond vao 0}
\E[\vanii_2 \cond \vao=0] & = \frac{1+p_1}{p_1+p_2}, \\
\label{eq: E vanii}
\E[\vanii_1] &= \E[\vanii_2] = \frac 1 {p_1+p_2}, \\
\label{eq: var vanii 1 - vanii 2 cond}
\Var[\vanii_1-\vanii_2 \cond \vao=1] & = \Var[\vanii_1-\vanii_2 \cond \vao=0] =
\frac{2(p_1(1-p_2)+p_2(1-p_1))}{(p_1+p_2)^2}.
\end{align}
\end{proposition}

Likewise, the OR $p_1(1-p_2)/(p_2(1-p_1))$ or the LOR $\log(p_1(1-p_2)/(p_2(1-p_1)))$ are equivalent to the odds $p/(1-p)$ or its logarithm if
\begin{equation}
\label{eq: p OR LOR}
p = \frac{p_1(1-p_2)}{p_1(1-p_2)+p_2(1-p_1)}.
\end{equation}
A Bernoulli random variable with parameter $p$ as in Eq.~\eqref{eq: p OR LOR} can be generated using a simpler algorithm than that used for RR and LRR.

\begin{algorithm}[Probability transformation for OR and LOR]
\label{algo: OR LOR, inner}
\ \\
\textbf{Inputs}: As many samples from sequences $\sei_1$, $\sei_2$ as needed.
\begin{enumerate}
\item
\label{step: inner OR LOR: go back}
Take a sample $\vai_1$ from sequence $\sei_1$ and a sample $\vai_2$ from sequence $\sei_2$.
\item
If $\vai_1 = \vai_2$, go to step~\ref{step: inner OR LOR: go back}. Else, set $\vao = 1$ if $\vai_1 = 1$ or $\vao = 0$ if $\vai_2 = 1$.
\end{enumerate}
\textbf{Output}: $\vao$.
\end{algorithm}

Algorithm~\ref{algo: OR LOR, inner} is similar to the method given by \citet{vonNeumann51} to generate a Bernoulli random variable with parameter $1/2$ from independent observations of one sequence $\sei_1$ with arbitrary $p_1$; in fact, it can be considered as an extension of that method to two populations. Von Neumann's procedure was refined by \citet{Elias72} and by \citet{Peres92}. Both authors presented methods that achieve an expected number of outputs per input arbitrarily close to the entropy of an input sample. It is not clear, however, if an analogous refinement exists in the two-population setting. In any case, for low $p_1$ and $p_2$ the estimation efficiency obtained with the proposed algorithm will be seen to be close to $1$, which suggests that there is little room for improvement, at least in that regime.

The numbers of input samples from $\sei_1$ and $\sei_2$ used by one execution of Algorithm~\ref{algo: OR LOR, inner} are again denoted by $\vanii_1$ and $\vanii_2$. In this case, by construction, $\vanii_1 = \vanii_2$.

\begin{proposition}
\label{prop: inner OR LOR}
Algorithm~\ref{algo: OR LOR, inner} terminates with probability $1$. Its output has $\Pr[\vao=1] = p$, $\Pr[\vao=0] = 1-p$, with $p$ given by Eq.~\eqref{eq: p OR LOR}. In addition, for $i=1,2$,
\begin{equation}
\label{eq: E vanii OR LOR}
\E[\vanii_i] = \E[\vanii_i \cond \vao=1] = \E[\vanii_i \cond \vao=0] = \frac 1 {p_1(1-p_2) + p_2(1-p_1)}.
\end{equation}
\end{proposition}

Repeatedly executing Algorithm~\ref{algo: RR LRR, inner}, or Algorithm~\ref{algo: OR LOR, inner}, produces a sequence $\seo$ of independent Bernoulli random variables with parameter $p$ given by Eq.~\eqref{eq: p RR LRR}, or by Eq.~\eqref{eq: p OR LOR}, to which the methods from \cite{Mendo25b} can be applied to obtain an unbiased estimator of either $p/(1-p)$ or $\log(p/(1-p))$; that is, of RR, LRR, OR or LOR. More specifically, let $\parg$ denote the parameter, of those four, that is to be estimated. Consider $\suc \in \mathbb N$, $\suc \geq 2$, and define
\begin{equation}
\label{eq: adjsuc}
\adjsuc =
\begin{cases}
1 & \text{for RR or OR}, \\
0 & \text{for LRR or LOR}.
\end{cases}
\end{equation}
The estimation method is then as follows.
First, an IBS procedure is applied, whereby samples of $\seo$ (generated from samples of $\sei_1$ and $\sei_2$) are consumed until $\suc+\adjsuc$ of them are \emph{successes}. Let $\vanoo'$ be the random number of samples of $\seo$ required for this. Then, a second IBS procedure is applied with ``failure'' and ``success'' swapped; that is, samples of $\seo$ are consumed until $\suc-\adjsuc$ of them are \emph{failures}, which requires a number $\vanoo''$ of samples of $\seo$. From $\vanoo'$ and $\vanoo''$, the estimation $\vahparg$ is computed as
\begin{align}
\label{eq: vahparg RR OR}
\vahparg &=
\frac{\suc\vanoo''}{(\suc-1) (\vanoo'-1)} \hspace{-28mm}& &\text{for RR or OR}, \\
\label{eq: vahparg LRR LOR}
\vahparg &=
-\harm{\vanoo'-1}+\harm{\vanoo''-1} \hspace{-28mm}& &\text{for LRR or LOR},
\end{align}
where $\harm{k}$ is defined in Section~\ref{part: intro}. This estimation
guarantees a certain accuracy by virtue of the following result, which stems directly from \cite[theorems 1 and 3]{Mendo25b}.

\begin{theorem}
\label{theo: MSE}
For $\suc \in \mathbb N$, $\suc \geq 2$, and for any $p_1, p_2 \in (0,1)$, the estimator for RR or OR defined by Eq.~\eqref{eq: vahparg RR OR} is unbiased, and
\begin{align}
\label{eq: rel MSE < RR OR}
\frac{\Var[\vahparg]}{\parg^2} &\leq \frac 1 {\suc-1} \left(1 - \frac{p(1-p)}{\suc-1+2p} \right) \\
\label{eq: rel MSE < const RR OR}
&< \frac 1 {\suc-1},
\end{align}
with $p$ given by Eq.~\eqref{eq: p RR LRR} for RR or Eq.~\eqref{eq: p OR LOR} for OR.

For $\suc \in \mathbb N$, $\suc \geq 2$, and for any $p_1, p_2 \in (0,1)$, the estimator for LRR or LOR defined by Eq.~\eqref{eq: vahparg LRR LOR} is unbiased, and
\begin{align}
\label{eq: MSE < LRR LOR}
\Var[\vahparg] &< \frac { \suc^2-\suc/4-1/4 } { (\suc-1+p)(\suc-p)(\suc-1/2)} - \frac{p(1-p)}{(\suc-1/2)^2} \left( 1 - \frac{1}{2\suc-3} \right) \\
\label{eq: MSE < const LRR LOR}
&<
\frac 1 {\suc-5/4},
\end{align}
with $p$ given by Eq.~\eqref{eq: p RR LRR} for LRR or by Eq.~\eqref{eq: p OR LOR} for LOR.
\end{theorem}

In view of Theorem~\ref{theo: MSE}, let $\cerr$ be defined as
\begin{equation}
\label{eq: cerr}
\cerr =
\begin{cases}
1 & \text{for RR or OR}, \\
5/4 & \text{for LRR or LOR}.
\end{cases}
\end{equation}
Then, for a target accuracy $\tarvar$, interpreted as relative MSE for RR or OR and as MSE for LRR or LOR, the parameter $\suc$ should be chosen as
\begin{equation}
\label{eq: suc tarvar}
\suc =
\left\lceil \frac 1 {\tarvar} + \cerr \right\rceil.
\end{equation}
This ensures that the accuracy is better than the target $\tarvar$ for any $p_1, p_2 \in (0,1)$.

Based on the above, the procedure for estimating RR or OR with guaranteed relative MSE, or for estimating LRR or LOR with guaranteed MSE, can be stated as follows.

\begin{algorithm}[Estimation of RR, LRR, OR or LOR]
\label{algo: outer}
\ \\
\textbf{Inputs}: Target relative MSE for RR or OR, or target MSE for LRR or LOR, denoted as $\tarvar$ in either case. As many samples from $\sei_1$, $\sei_2$ as needed.
\begin{enumerate}
\item
Compute $\adjsuc$ and $\cerr$ from Eqs.~\eqref{eq: adjsuc} and \eqref{eq: cerr}.
\item
Compute $\suc$ from Eq.~\eqref{eq: suc tarvar}.
\item
\label{step: outer: IBS 1}
Generate samples from $\seo$ until exactly $\suc+\adjsuc$ of them are successes, which requires a total of $\vanoo'$ samples from $\seo$. Each of these samples is generated using Algorithm~\ref{algo: RR LRR, inner} for RR or LRR, or  Algorithm~\ref{algo: OR LOR, inner} for OR or LOR, taking samples from $\sei_1$ and $\sei_2$ as inputs.
\item
\label{step: outer: IBS 2}
Generate samples from $\seo$ until exactly $\suc-\adjsuc$ of them are failures, which requires a total of $\vanoo''$ samples from $\seo$. These samples are generated as in the previous step.
\item
Compute the estimation $\vahparg$ using Eq.~\eqref{eq: vahparg RR OR} for RR or OR, or Eq.~\eqref{eq: vahparg LRR LOR} for LRR or LOR.
\end{enumerate}
\textbf{Output}: $\vahparg$.
\end{algorithm}

The estimation procedure described in Algorithm~\ref{algo: outer} consists of an \emph{outer loop} with two IBS procedures applied on samples of $\seo$, and an \emph{inner loop} that generates those samples using observations from $\sei_1$ and $\sei_2$. The outer loop is the same for RR, LRR, OR or LOR estimation (only with different values of $\adjsuc$ and $\cerr$), but the inner loop is different for RR or LRR on one hand, and for OR or LOR on the other hand (corresponding to Algorithms~\ref{algo: RR LRR, inner} and \ref{algo: OR LOR, inner} respectively).

Algorithm~\ref{algo: outer} has been formulated considering that observations from either $\sei_1$ or $\sei_2$ may be taken \emph{as needed}. Let $\vanoi_1$ and $\vanoi_2$ respectively denote the total numbers of observations from $\sei_1$ and $\sei_2$ that are used by the algorithm. For RR and LRR estimation, $\vanoi_1$ and $\vanoi_2$ are not necessarily equal, because each run of Algorithm~\ref{algo: RR LRR, inner} may not take the same number of inputs from the two populations. On the other hand, for OR and LOR it is always the case that $\vanoi_1 = \vanoi_2$, because Algorithm~\ref{algo: OR LOR, inner} takes its inputs in pairs.

From the preceding paragraph it is seen that, in general, the total numbers of observations from the two populations required by the estimator, i.e.~$\vanoi_1$ and $\vanoi_2$, may not be equal. However, by assumption, the observations from $\sei_1$ and $\sei_2$ are taken \emph{in pairs} of one sample from each population. The way to reconcile these two standpoints is to take the samples in pairs in a ``conservative'' way, as in \cite{Mendo25c}: whenever it is necessary to take a pair of samples, namely because a sample from either the first or the second population is needed by the estimation procedure, the sample from the other population is \emph{stored for later use}; and a new pair will subsequently be taken \emph{only if necessary}, i.e.~if a sample is needed from a population for which no surplus samples are available from previous pairs. Samples remaining at the end of the process, if any, are discarded. By this procedure, the number $\vanoi$ of required pairs is
\begin{equation}
\label{eq: vanoi max}
\vanoi = \max\{\vanoi_1, \vanoi_2\}.
\end{equation}

The sampling procedure that has been described, represented by Eq.~\eqref{eq: vanoi max}, incurs some loss of efficiency for RR and LRR, with some samples discarded at the end of the estimation process unless $\vanoi_1$ and $\vanoi_2$ happen to be equal. For OR and LOR there is no such loss, because $\vanoi_1$ and $\vanoi_2$ are necessarily equal. In general, noting that
\begin{equation}
\label{eq: max suma dif abs}
\max\{\vanoi_1, \vanoi_2\} = \frac{\vanoi_1 + \vanoi_2} 2 + \frac{|\vanoi_1 - \vanoi_2|} 2,
\end{equation}
it is clear that for a specific value of $\vanoi_1+\vanoi_2$ the number of pairs, $\vanoi$, must be at least $(\vanoi_1+\vanoi_2)/2$, and this bound is achieved when $\vanoi_1=\vanoi_2$. Thus, a \emph{sampling efficiency factor} can be defined as
\begin{equation}
\label{eq: sef}
\sef = \frac{\E[\vanoi_1+\vanoi_2]}{2\E[\vanoi]}.
\end{equation}
It follows from Eqs.~\eqref{eq: vanoi max}--\eqref{eq: sef} 
that $1/2 \leq \sef \leq 1$, and that $\sef$ is close to $1$ if $\E[|\vanoi_1-\vanoi_2|]$ is small compared with $\E[\vanoi_1+\vanoi_2]$.

The next section characterizes $\E[\vanoi]$ and $\sef$. For RR and LRR this involves obtaining the joint distribution of $\vanoi_1$ and $\vanoi_2$.

\section{Average number of input pairs}
\label{part: E max, gen}

\subsection{For relative risk and log relative risk}
\label{part: E max, RR LRR}

The variables $\vanoi_1$ and $\vanoi_2$ for RR and LRR are statistically dependent, for the same reason $\vanii_1$ and $\vanii_2$ are. The following proposition characterizes their joint distribution.

\begin{proposition}
\label{prop: outer RR LRR}
For RR or LRR estimation, the joint probability function of the numbers of inputs used by Algorithm~\ref{algo: outer}, $\vanoi_1$ and $\vanoi_2$, is
\begin{equation}
\label{eq: Pr vanoi 1 vanoi 2 RR LRR}
\begin{split}
&\Pr[\vanoi_1 = \noi_1, \vanoi_2 = \noi_2] = \\
&\ \frac{(1-p_1)^{\noi_1}(1-p_2)^{\noi_2}}{2^{\noi_1+\noi_2}}
\sum_{\noo'=\suc+\adjsuc}^{\noi_2+2\adjsuc} \, \sum_{\noo''=\suc-\adjsuc}^{\noi_1-2\adjsuc}
\binom{\noo'-1}{\suc+\adjsuc-1}
\binom{\noo''-1}{\suc-\adjsuc-1} \\
&\ \cdot \binom{\noi_1+\noi_2-1}{\noi_1-\noo''-2\adjsuc,\, \noi_2-\noo'+2\adjsuc,\, \noo'+\noo''-1} \left(\frac{p_1}{1-p_1}\right)^{\noo''+2\adjsuc} \left(\frac{p_2}{1-p_2}\right)^{\noo'-2\adjsuc},
\end{split}
\end{equation}
with $\noi_1 \geq \suc+\adjsuc$, $\noi_2 \geq \suc-\adjsuc$.
In addition,
\begin{equation}
\label{eq: vanoi 1 RR LRR}
\E[\vanoi_1] = \E[\vanoi_2] = \frac{\suc+\adjsuc}{p_1} + \frac{\suc-\adjsuc}{p_2}.
\end{equation}
\end{proposition}

Using Proposition~\ref{prop: outer RR LRR}, the average number of required input pairs for RR and LRR can be computed as 
\begin{equation}
\label{eq: E vanoi RR LRR}
\begin{split}
\E[\vanoi] &=
\sum_{\noi_1=\suc+\adjsuc}^\infty \, \sum_{\noi_2=\suc-\adjsuc}^\infty \max\{\noi_1, \noi_2\} \Pr[\vanoi_1 = \noi_1, \vanoi_2 = \noi_2] \\
&=
\sum_{\noi = \suc+\adjsuc}^\infty \noi \left( \sum_{\noi_1=\suc+\adjsuc}^{\noi} \Pr[\vanoi_1 = \noi_1, \vanoi_2 = \noi] + \sum_{\noi_2=\suc-\adjsuc}^{\noi-1} \Pr[\vanoi_1 = \noi, \vanoi_2 = \noi_2] \right).
\end{split}
\end{equation}
Obtaining $\E[\vanoi]$ from Eq.~\eqref{eq: E vanoi RR LRR} is computationally intensive, particularly for small values of $p_1$ and $p_2$, as these result in slow convergence of the series. A lower bound, which also provides a good approximation, is given next. Let
\begin{align}
\label{eq: roprpr}
\roprpr &= \sqrt{p_1 p_2}, \\
\label{eq: RR}
\RR &= \frac{p_1}{p_2}.
\end{align}

\begin{proposition}
\label{prop: E max cota RR LRR}
For RR or LRR estimation, the required number of input pairs $\vanoi$
and the sampling efficiency factor defined by Eq.~\eqref{eq: sef} satisfy the following:
\begin{align}
\label{eq: E vanoi < RR LRR}
\E[\vanoi] &< \frac{\suc+\adjsuc}{p_1} + \frac{\suc-\adjsuc}{p_2} + \sqrt{\frac{\suc+\adjsuc}{2p_1} + \frac{\suc-\adjsuc}{2p_2}}
, \\
\label{eq: sef >}
\sef &> \frac 1 {1 + \sqrt{\displaystyle \frac{p_1 p_2}{2\left( (\suc+\adjsuc)p_2+(\suc-\adjsuc)p_1 \right)}}}
= \frac 1 {1 + \sqrt{ \displaystyle \frac {\roprpr} {2\left( (\suc+\adjsuc)/\sqrt{\RR} + (\suc-\adjsuc)\sqrt{\RR} \right)} } }, \\
\label{eq: lim sef}
\lim_{\roprpr \rightarrow 0} \sef &= 1.
\end{align}
\end{proposition}

\begin{figure}%
\centering%
\begin{subfigure}{\tamdosfig}%
\centering%
\includegraphics[width=\textwidth]{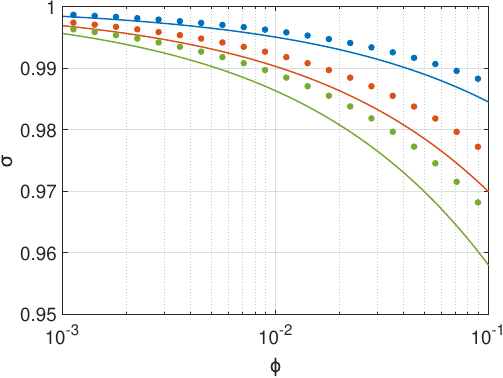}%
\caption{$\RR = 1$}%
\label{fig: sef RR 1}%
\end{subfigure}%
\hfill%
\begin{subfigure}{\tamdosfig}%
\centering%
\includegraphics[width=\textwidth]{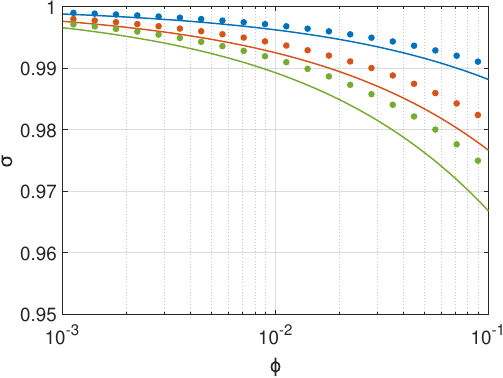}%
\caption{$\RR = 10$}%
\label{fig: sef RR 10}%
\end{subfigure}%
\\%
\begin{subfigure}{\tamdosfig}%
\centering%
\includegraphics[width=\textwidth]{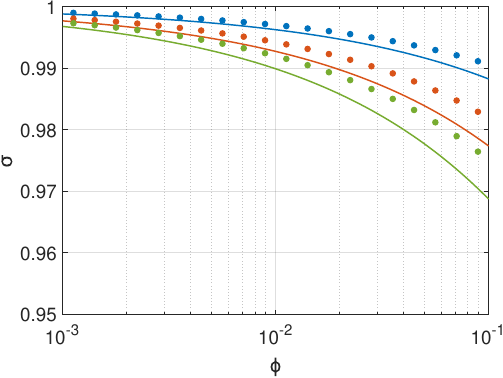}%
\caption{$\RR = 0.1$}%
\label{fig: sef RR 0p1}%
\end{subfigure}%
\hfill%
\begin{subfigure}{\tamdosfig}%
\centering%
\includegraphics[scale=\legendscale]{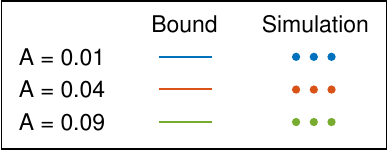}%
\hspace{-5mm}\vspace*{20mm}
\caption*{}%
\end{subfigure}%
\caption{Sampling efficiency factor for RR estimation}%
\label{fig: sef RR}%
\end{figure}

Figure~\ref{fig: sef RR} shows Monte Carlo simulation results of the sampling efficiency factor $\sef$ for the RR estimator. Each simulation consists of $10^7$ realizations of the estimation procedure, from which $\sef$ is obtained using Eq.~\eqref{eq: sef} with expected values replaced by sample means. The bound given by Proposition~\ref{prop: E max cota RR LRR} is also plotted. In this and subsequent figures, the parameters $\roprpr$ and $\RR$ are used rather than $p_1$ and $p_2$ for convenience, and the following range of values is considered: $\roprpr$ between $0.001$ and $0.1$; $\RR= 1$, $10$ and $0.1$; $\tarvar= 0.01$, $0.04$ and $0.09$ (corresponding to a relative RMSE or RMSE target equal to $10\%$, $20\%$ and $30\%$). These values span a wide gamut of typical conditions in practical use cases. As seen in Figure~\ref{fig: sef RR}, $\sef$ is very close to $1$ for the considered values of $\roprpr$, $\RR$, $\tarvar$; that is, the efficiency loss caused by sampling in pairs is small. The figure also shows that the bound is a good approximation to the actual $\sef$. Results for the LRR estimator are similar, and are omitted.

\subsection{For odds ratio and log odds ratio}
\label{part: E max, OR LOR}

When estimating OR or LOR, since Algorithm~\ref{algo: OR LOR, inner} is employed as inner loop, the total numbers of inputs $\vanoi_1$ and $\vanoi_2$ used by Algorithm~\ref{algo: outer} are equal. Thus $\E[\vanoi] = \E[\vanoi_1] = \E[\vanoi_2]$, and $\sef=1$. Computing $\E[\vanoi_i]$, $i=1,2$ is also easy in this case because, according to Proposition~\ref{prop: inner OR LOR}, the conditional mean of $\vanii_i$ does not depend on $\vao$.

\begin{proposition}
\label{prop: outer OR LOR}
For OR and LOR estimation, the numbers of inputs used by Algorithm~\ref{algo: outer}, $\vanoi_1$ and $\vanoi_2$, have the following mean:
\begin{equation}
\label{eq: vanoi OR LOR}
\E[\vanoi_1] = \E[\vanoi_2]
= \frac{\suc+\adjsuc}{p_1(1-p_2)} + \frac{\suc-\adjsuc}{p_2(1-p_1)}.
\end{equation}
\end{proposition}

\section{Estimation efficiency}
\label{part: effic}

The \emph{efficiency} of the proposed sequential estimators can be defined, as argued in \cite{Mendo25c}, by comparing the estimation variance with the smallest variance that can be achieved by a fixed-size estimator with the same average number of samples from each population, which is given by the vector form of the \CR{} bound \cite[chapter~3]{Kay93}. Consider an arbitrary parameter $\parg$ that is a differentiable function of $p_1$ and $p_2$. For an unbiased estimator $\vahparg$ of $\parg$, which uses independent observations of the two populations taken in pairs, the referred bound gives
\begin{equation}
\label{eq: efic gen}
\effic = \frac{\left(\displaystyle \frac{\partial \parg}{\partial p_1}\right)^2 p_1(1-p_1) + \left(\displaystyle \frac{\partial \parg}{\partial p_2}\right)^2 p_2(1-p_2)}{\E[\vanoi] \Var[\vahparg]},
\end{equation}
where $\vanoi$ is the number of pairs. From this expression, the efficiency of the considered estimators can be characterized as given next. Based on Theorem~\ref{theo: MSE}, let $\vef$ be defined as either Eq.~\eqref{eq: rel MSE < RR OR} divided by Eq.~\eqref{eq: rel MSE < const RR OR} or Eq.~\eqref{eq: MSE < LRR LOR} divided by Eq.~\eqref{eq: MSE < const LRR LOR}, depending on the estimator:
\begin{equation}
\label{eq: vef}
\vef = \begin{cases}
1 - \displaystyle \frac{p(1-p)}{\suc-1+2p} & \text{for RR and OR}, \\[4mm] 
\displaystyle \frac { (\suc^2-\suc/4-1/4)(\suc-5/4) } { (\suc-1+p)(\suc-p)(\suc-1/2)} - \frac{p(1-p)(\suc-2)(\suc-5/4)}{(\suc-1/2)^2(\suc-3/2)}  & \text{for LRR and LOR},
\end{cases}
\end{equation}
with $p$ given by Eq.~\eqref{eq: p RR LRR} for RR and LRR, or by Eq.~\eqref{eq: p OR LOR} for OR and LOR.
It follows from Theorem~\ref{theo: MSE} that $\vef<1$.

\begin{theorem}
\label{theo: effic RR LRR OR LOR}
The efficiency of the RR and LRR estimators is bounded for any $p_1, p_2 \in (0,1)$ as
\begin{equation}
\label{eq: efic RR LRR}
\begin{split}
\effic &>
\frac{\left(p_2(1-p_1) + p_1(1-p_2)\right) (\suc-\cerr)} {(\suc+\adjsuc)p_2 + (\suc-\adjsuc)p_1} \frac{\sef}{\vef} \\[0.5mm] 
&= \frac{\left(\displaystyle\frac 1 {\sqrt{\RR}}+\sqrt{\RR}-2\roprpr\right) (\suc-\cerr)}{\displaystyle\frac{\suc+\adjsuc}{\sqrt{\RR}} + (\suc-\adjsuc)\sqrt{\RR}} \frac{\sef}{\vef},
\end{split}
\end{equation}
where $\sef$ is defined by Eq.~\eqref{eq: sef} and bounded by Proposition~\ref{prop: E max cota RR LRR}, and $\cerr$ and $\vef$ are given by Eqs.~\eqref{eq: cerr} and \eqref{eq: vef}.

The efficiency of the OR and LOR estimators is bounded for any $p_1, p_2 \in (0,1)$ as
\begin{equation}
\label{eq: efic OR LOR}
\begin{split}
\effic &>
\frac{\left(p_1(1-p_1) + p_2(1-p_2)\right) (\suc-\cerr)} {(\suc+\adjsuc)p_2(1-p_1) + (\suc-\adjsuc)p_1(1-p_2)} \frac{1}{\vef} \\[0.5mm] 
&= \frac{\left(\displaystyle \frac 1 {\sqrt{\RR}} + \sqrt{\RR} - \roprpr \left(\displaystyle \frac 1 {\RR} + \RR \right) \right) (\suc-\cerr)} {(\suc+\adjsuc)\left(\displaystyle \frac 1 {\sqrt{\RR}} - \roprpr\right) + (\suc-\adjsuc)\left(\sqrt{\RR}-\roprpr\right)} \frac{1}{\vef},
\end{split}
\end{equation}
where $\cerr$ and $\vef$ are given by Eqs.~\eqref{eq: cerr} and \eqref{eq: vef}.
\end{theorem}

\begin{figure}%
\centering%
\begin{subfigure}{\tamdosfig}%
\centering%
\includegraphics[width=\textwidth]{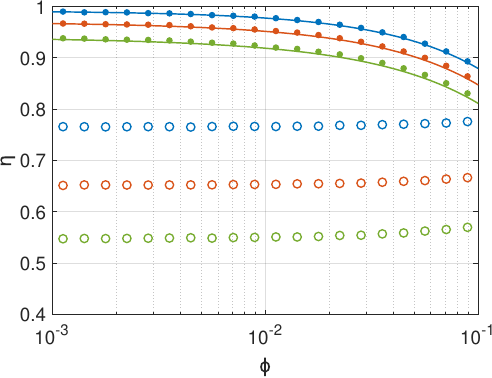}%
\caption{$\RR = 1$}%
\label{fig: effic RR 1}%
\end{subfigure}%
\hfill%
\begin{subfigure}{\tamdosfig}%
\centering%
\includegraphics[width=\textwidth]{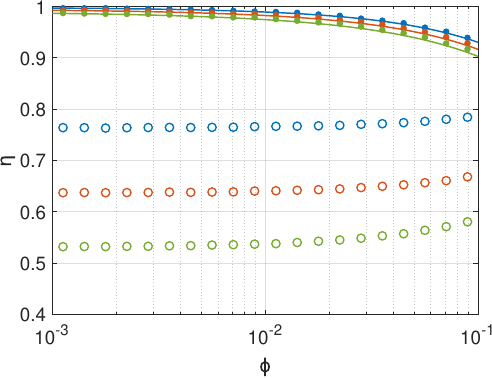}%
\caption{$\RR = 10$}%
\label{fig: effic RR 10}%
\end{subfigure}%
\\%
\begin{subfigure}{\tamdosfig}%
\centering%
\includegraphics[width=\textwidth]{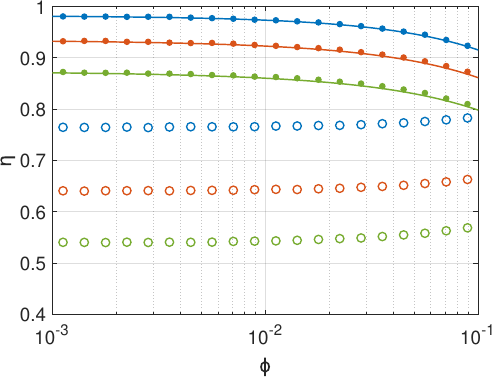}%
\caption{$\RR = 0.1$}%
\label{fig: effic RR 0p1}%
\end{subfigure}%
\hfill%
\begin{subfigure}{\tamdosfig}%
\centering%
\includegraphics[scale=\legendscale]{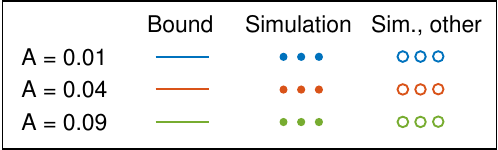}%
\hspace{-3mm}\vspace*{20.2mm}
\caption*{}%
\end{subfigure}%
\caption{Estimation efficiency for RR}%
\label{fig: effic RR}%
\end{figure}%

Figure~\ref{fig: effic RR} shows simulation results for the efficiency of the RR estimator. The simulation is similar to that described in Subsection~\ref{part: E max, RR LRR}: for each combination of input parameters $\roprpr$, $\RR$ and $\tarvar$, $10^7$ realizations of the estimator are simulated. The efficiency is computed using Eq.~\eqref{eq: efic gen} particularized for $\parg = \RR$, with $\E[\vanoi]$ replaced by the average number of required pairs resulting from the simulation and $\Var[\vahparg]$ replaced by the sample MSE. The same range of values for $\roprpr$, $\RR$ and $\tarvar$ as in Subsection~\ref{part: E max, RR LRR} is used. The bound given by Theorem~\ref{theo: effic RR LRR OR LOR} is also plotted. In addition, for comparison purposes, simulation results are shown for the group-sampling version of the estimation method described in \cite{Mendo25c}, particularized to groups of one sample from each population (i.e.~$l_1 = l_2 = 1$ in that reference). 

As seen in the figure, the efficiency increases as $\roprpr$ becomes smaller, and for $\roprpr$ moderately small it already reaches values near $1$, well above the efficiency of the reference method. The theoretical bound is seen to be a very good approximation to the actual values obtained from simulation. It is also observed that the efficiency is higher when the target accuracy $\tarvar$ is smaller, i.e.~more demanding. This happens because the terms $\suc+\adjsuc$, $\suc-\adjsuc$ and $\suc-\cerr$ in Eq.~\eqref{eq: efic RR LRR} from Theorem~\ref{theo: effic RR LRR OR LOR} become relatively more similar to each other as $\tarvar$ is reduced, or equivalently as $\suc$ increases. Lastly, the efficiency is better for $\RR$ large than for $\RR$ small, and this effect is more noticeable when $\tarvar$ is larger. This is due to the fact that for large $\RR$ the summand containing the factor $\suc-\adjsuc$ dominates in the denominator of Eq.~\eqref{eq: efic RR LRR}, whereas for small $\RR$ the summand with $\suc+\adjsuc$ is the dominant one.

\begin{figure}%
\centering%
\begin{subfigure}{\tamdosfig}%
\centering%
\includegraphics[width=\textwidth]{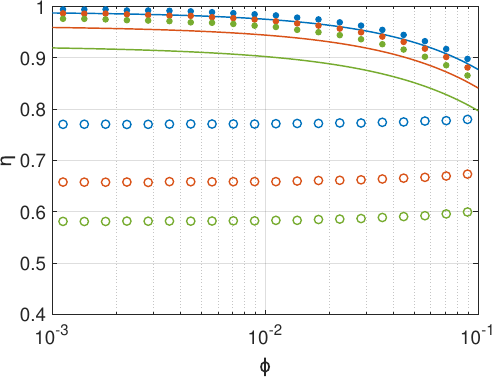}%
\caption{$\RR = 1$}%
\label{fig: effic LRR 1}%
\end{subfigure}%
\hfill%
\begin{subfigure}{\tamdosfig}%
\centering%
\includegraphics[width=\textwidth]{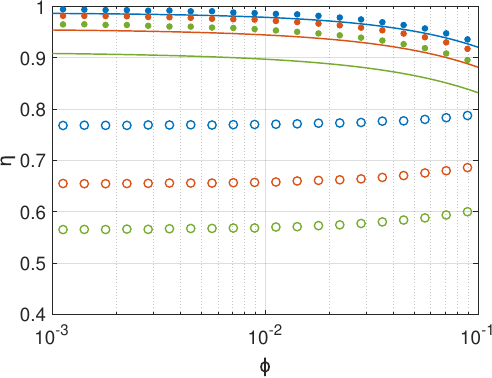}%
\caption{$\RR = 10$}%
\label{fig: effic LRR 10}%
\end{subfigure}%
\\%
\begin{subfigure}{\tamdosfig}%
\centering%
\includegraphics[scale=\legendscale]{legend_effic}%
\caption*{\vspace{-4mm}}%
\end{subfigure}%
\caption{Estimation efficiency for LRR}%
\label{fig: effic LRR}%
\end{figure}%

For LRR, the efficiency of the proposed estimator is unchanged upon replacing $\RR$ by $1/\RR$, as is justified next. For a given $\roprpr$, inverting $\RR$ is equivalent to interchanging $p_1$ and $p_2$, or replacing $p$ by $1-p$. The fact that $\adjsuc$ is $0$ for LRR implies that the error variance $\Var[\vahparg]$ is symmetric with respect to those changes, as are the expressions of  $\E[\vanoi_1]$, $\E[\vanoi_2]$, $\E[\vanoi]$ and $\vef$, and therefore also those of $\sef$ and $\effic$. The method from \cite{Mendo25c} used for comparison also has this symmetry for LRR when sampling is done in groups of one sample from each population. These observations also apply to LOR.

Figure~\ref{fig: effic LRR} presents the results for LRR. By the argument in the preceding paragraph, the efficiency values for $\RR = 10$ and for $\RR= 0.1$ are necessarily equal (up to the random fluctuations inherent to Monte Carlo simulation), and therefore the graphs for $\RR = 0.1$ are omitted. The general trends observed in Figure~\ref{fig: effic LRR} are similar to those noticed for RR (Figure~\ref{fig: effic RR}), except for two differences. Firstly, for LRR the efficiency obtained from simulation is less sensitive to $\RR$ (and is known to be the same when $\RR$ is replaced by $1/\RR$). This also applies to the bound given by Theorem~\ref{theo: effic RR LRR OR LOR}: the fact that $\adjsuc$ is $0$ for LRR diminishes the influence of $\RR$ on the right-hand side of Eq.~\eqref{eq: efic RR LRR}. Secondly, this bound is less tight than it was for RR. The reason is that the bound for LRR is based on that obtained in \cite{Mendo25b} for log odds estimation, which is not very tight, as discussed in that reference, due to the difficulty of dealing with the logarithm function.

\begin{figure}%
\centering%
\begin{subfigure}{\tamdosfig}%
\centering%
\includegraphics[width=\textwidth]{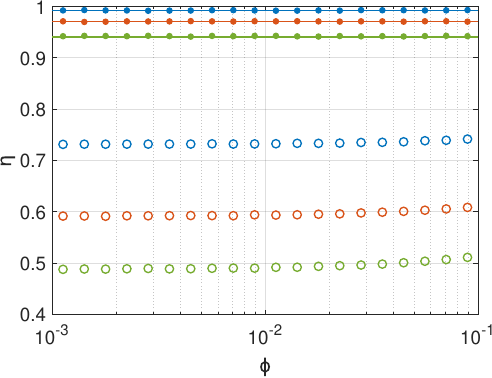}%
\caption{$\RR = 1$}%
\label{fig: effic OR 1}%
\end{subfigure}%
\hfill%
\begin{subfigure}{\tamdosfig}%
\centering%
\includegraphics[width=\textwidth]{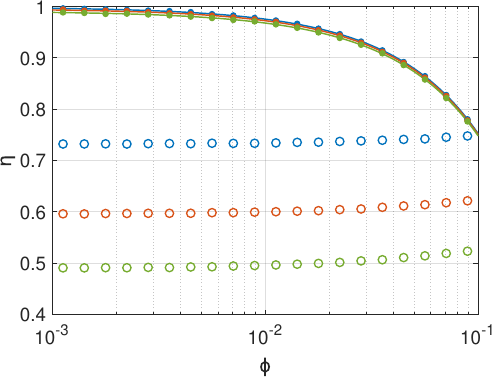}%
\caption{$\RR = 10$}%
\label{fig: effic OR 10}%
\end{subfigure}%
\\%
\begin{subfigure}{\tamdosfig}%
\centering%
\includegraphics[width=\textwidth]{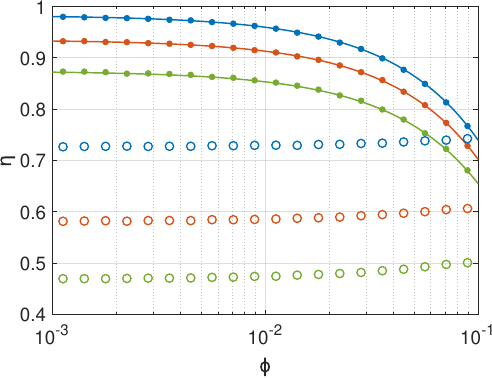}%
\caption{$\RR = 0.1$}%
\label{fig: effic OR 0p1}%
\end{subfigure}%
\hfill%
\begin{subfigure}{\tamdosfig}%
\centering%
\includegraphics[scale=\legendscale]{legend_effic}%
\hspace{-3mm}\vspace*{20.2mm}
\caption*{}%
\end{subfigure}%
\caption{Estimation efficiency for OR}%
\label{fig: effic OR}%
\end{figure}%

\begin{figure}%
\centering%
\begin{subfigure}{\tamdosfig}%
\centering%
\includegraphics[width=\textwidth]{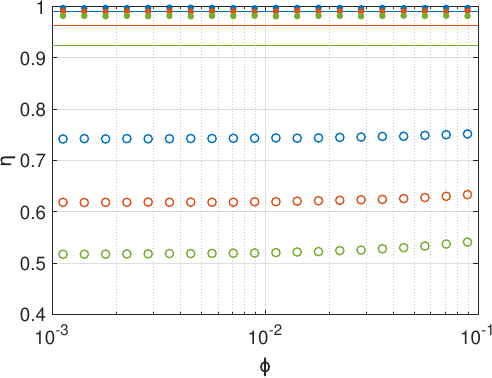}%
\caption{$\RR = 1$}%
\label{fig: effic LOR 1}%
\end{subfigure}%
\hfill%
\begin{subfigure}{\tamdosfig}%
\centering%
\includegraphics[width=\textwidth]{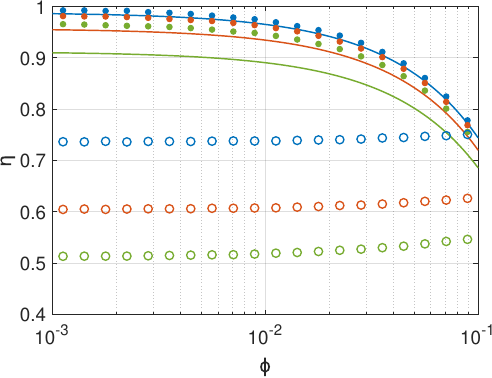}%
\caption{$\RR = 10$}%
\label{fig: effic LOR 10}%
\end{subfigure}%
\\%
\begin{subfigure}{\tamdosfig}%
\centering%
\includegraphics[scale=\legendscale]{legend_effic}%
\caption*{\vspace{-4mm}}%
\end{subfigure}%
\caption{Estimation efficiency for LOR}%
\label{fig: effic LOR}%
\end{figure}%

The efficiency results for OR and LOR are plotted in Figures~\ref{fig: effic OR} and \ref{fig: effic LOR}. Similar observations can be made as for RR and LRR: $\effic$ tends to increase as $\roprpr$ or $\tarvar$ decrease, and is better than that of the reference method for small or moderately small $\roprpr$; the theoretical bound is tighter for OR than for LOR; and for OR the efficiency is better for $\RR$ large than for $\RR$ small. A difference with respect to RR and LRR is that for OR and LOR the efficiency of the proposed estimator is independent of $\roprpr$ when $\RR = 1$. Indeed, in Eq.~\eqref{eq: efic OR LOR} from Theorem~\ref{theo: effic RR LRR OR LOR} the fraction with the explicit dependence on $\roprpr$ is seen to become independent of this parameter when $\RR=1$, and the term $\vef$ is also independent of $\roprpr$ when $\RR=1$ because $p$ is.

It is of interest to characterize the efficiency of the estimators for unknown $\RR$; that is, to obtain a bound on $\effic$ that is \emph{independent of $\RR$}. This is addressed in what follows. An important particular case for practical applications is the \emph{small-probability regime}, whereby the considered attribute is rare in the observed populations. In that case, $\RR$ is unknown but $p_1$ and $p_2$ can be assumed to take small values.

For the subsequent analysis, it will be useful to define
\begin{equation}\
\label{eq: prmax}
\prmax = \max\{p_1, p_2\} = \roprpr \max\left\{\sqrt{\RR}, 1/\sqrt{\RR}\right\},
\end{equation}
which implies that
\begin{equation}
\label{eq: roprpr prmax}
\roprpr = \prmax \min\left\{\sqrt{\RR}, 1/\sqrt{\RR}\right\}.
\end{equation}
In addition, the simple fact stated by the proposition below will be needed.

\begin{proposition}
\label{prop: RR roprpr prmaxsup}
Assume that $\prmax$ is less than some $\prmaxsup \leq 1$. Then, $\roprpr$ must be less than $\prmaxsup \min\{\sqrt{\RR}, 1/\sqrt{\RR}\}$; and given any such $\roprpr$, $\RR$ must be in the interval $(\roprpr^2/\prmaxsup^2,\, \prmaxsup^2/\roprpr^2)$.
\end{proposition}

For RR and LRR, a bound independent of $\RR$ is given by the next theorem.

\begin{theorem}
\label{theo: effic RR LRR, indep of RR}
The efficiency of the RR and LRR estimators is bounded for $p_1, p_2 \in (0,1)$ as
\begin{equation}
\label{eq: efic RR LRR, indep of RR}
\effic > \frac{\suc-\cerr}{\suc+\adjsuc} \frac{\suc-\sqrt{\adjsuc^2+\roprpr^2(\suc^2-\adjsuc^2)}}{\suc-\adjsuc} \frac {2\sqrt[4]{\suc^2-\adjsuc^2}} {2\sqrt[4]{\suc^2-\adjsuc^2} + \sqrt{\roprpr}},
\end{equation}
where $\roprpr$ is defined in Eq.~\eqref{eq: roprpr}. This bound is a decreasing function of $\roprpr$, and
\begin{equation}
\label{eq: efic RR LRR, indep of RR, lim}
\liminf_{\roprpr \rightarrow 0} \effic \geq \frac{\suc-\cerr}{\suc+\adjsuc} \geq \frac 1 {1 + \tarvar(\cerr+\adjsuc)}.
\end{equation}
\end{theorem}

\begin{figure}%
\centering%
\begin{subfigure}{\tamdosfig}%
\centering%
\includegraphics[width=\textwidth]{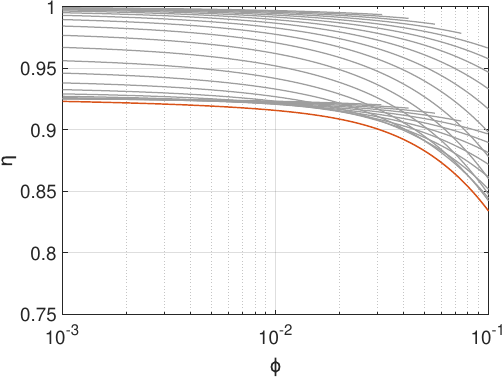}%
\caption{RR}%
\label{fig: effic RR bounds 0p04}%
\end{subfigure}%
\hfill%
\begin{subfigure}{\tamdosfig}%
\centering%
\includegraphics[width=\textwidth]{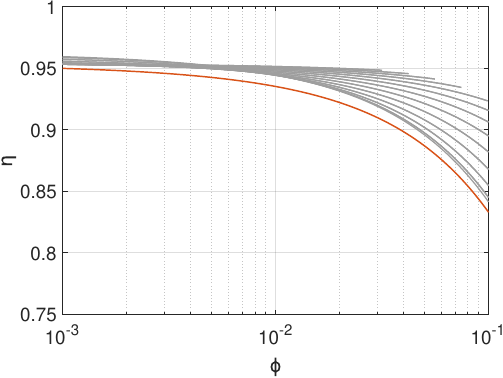}%
\caption{LRR}%
\label{fig: effic LRR bounds 0p04}%
\end{subfigure}%
\\%
\begin{subfigure}{\tamdosfig}%
\centering%
\includegraphics[scale=\legendscale]{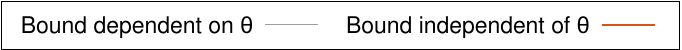}%
\caption*{\vspace{-4mm}}%
\end{subfigure}%
\caption{Bounds on estimation efficiency for RR and LRR. Example for $\tarvar=0.04$}%
\label{fig: effic RR LRR bounds 0p04}%
\end{figure}%

A comparison can be seen in Figure~\ref{fig: effic RR LRR bounds 0p04}, for RR and LRR and using $\tarvar= 0.04$ as an example, between the bound in Theorem~\ref{theo: effic RR LRR, indep of RR} (thick, red curve) and that in Theorem~\ref{theo: effic RR LRR OR LOR}, which depends on $\RR$. The latter bound is plotted for $25$ values of $\RR$ logarithmically spaced from $0.001$ to $1000$ (thin, grey lines). Due to the symmetry in LRR discussed previously, not all values of $\RR$ produce a distinct curve in Figure~\ref{fig: effic LRR bounds 0p04}. Note also that, according to Eq.~\eqref{eq: roprpr prmax}, for a given $\RR$ it is not possible for $\roprpr$ to exceed $\min\{\sqrt{\RR}, 1/\sqrt{\RR}\}$. This is the reason
some of the curves for specific values of $\RR$ do not span the full range of $\roprpr$ shown in the graph. It can be observed that among all the $\RR$-specific curves, the lowest one is not the same for all $\roprpr$; that is, the worst-case value of $\RR$ in Theorem~\ref{theo: effic RR LRR OR LOR}
depends on $\roprpr$. The bound from Theorem~\ref{theo: effic RR LRR, indep of RR} is seen to be below all these curves, and close to the lowest one for each $\roprpr$.

\begin{figure}%
\centering%
\begin{subfigure}{\tamdosfig}%
\centering%
\includegraphics[width=\textwidth]{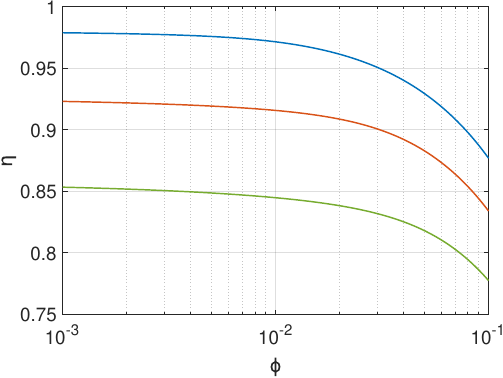}%
\caption{RR}%
\label{fig: effic RR bound}%
\end{subfigure}%
\hfill%
\begin{subfigure}{\tamdosfig}%
\centering%
\includegraphics[width=\textwidth]{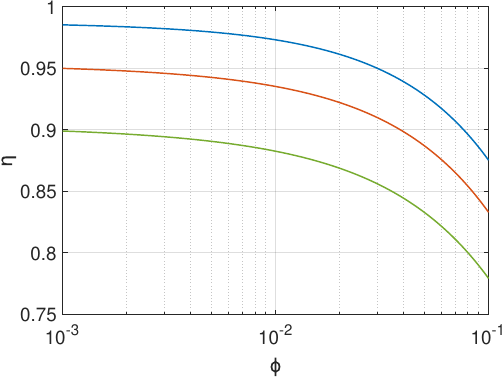}%
\caption{LRR}%
\label{fig: effic LRR bound}%
\end{subfigure}%
\\%
\begin{subfigure}{\tamdosfig}%
\centering%
\includegraphics[scale=\legendscale]{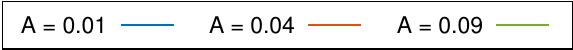}
\caption*{\vspace{-4mm}}%
\end{subfigure}%
\caption{Bound on estimation efficiency independent of $\RR$, for RR and LRR}%
\label{fig: effic RR LRR bound}%
\end{figure}%

Figure~\ref{fig: effic RR LRR bound} shows the bound on the estimation efficiency for RR and LRR given by Theorem~\ref{theo: effic RR LRR, indep of RR}, for the same values of $\tarvar$ as in Figures~\ref{fig: effic RR}--\ref{fig: effic LOR}. 
The fact that this bound decreases with $\roprpr$ implies that, if $\roprpr$ is assumed not to exceed a given value $\roprprsup$, the efficiency will be higher than the bound particularized to $\roprprsup$. As an example, for $\tarvar=0.04$, if $\roprpr \leq 10^{-2}$ (i.e.~if $\sqrt{p_1 p_2} \leq 10^{-2}$, or in particular if $p_1+p_2 \leq 0.02$) the efficiency of the RR and LRR estimators is guaranteed to be better than $91.5\%$ and $93.5\%$ respectively, regardless of $\RR$. This means that, to achieve the same accuracy, the number of input pairs used by the best fixed-size estimator would be at least $0.915$ or $0.935$ times the average number of pairs used by the proposed estimator. For $\tarvar = 0.04$, as $\roprpr \rightarrow 0$ the efficiency takes values above $92.5\%$ and $95.3\%$ for RR and LRR respectively.

The behaviour of the estimation efficiency for OR and LOR is slightly different from that for RR and LRR, as explained next. According to Proposition~\ref{prop: RR roprpr prmaxsup}, for a given $\roprpr$ and without any restriction on $\prmax$, i.e.~$\prmaxsup=1$, it is possible for $\RR$ to take any value in the interval $(\roprpr^2,\, 1/\roprpr^2)$. For OR and LOR, although the $\RR$-dependent bound in Theorem~\ref{theo: effic RR LRR OR LOR} converges to a positive value as $\roprpr \rightarrow 0$ \emph{with $\RR$ fixed}, it becomes arbitrarily close to $0$ if $\roprpr$ is sufficiently small and $\RR$ is sufficiently close to either $\roprpr^2$ or $1/\roprpr^2$. Thus, unlike what happened for RR and LRR, restricting $\roprpr$ to be less than or equal to a given $\roprprsup$, with $\RR$ arbitrary, is not enough to produce a positive lower bound independent of $\RR$.

The underlying reason for this different behaviour is that the bound for RR and LRR in Theorem~\ref{theo: effic RR LRR OR LOR} only becomes small if \emph{both} $p_1$ and $p_2$ are large, and that possibility is excluded by making $\roprpr$ small. On the other hand, the bound for OR and LOR becomes small if \emph{one} of those probabilities is large while the other is small, and this can happen even for small $\roprpr$ provided that $\RR$ is close to $\roprpr^2$ or $1/\roprpr^2$. Nevertheless, restricting $\prmax$ to be less than a given $\prmaxsup < 1$ prevents this, because then Proposition~\ref{prop: RR roprpr prmaxsup} implies that $\RR$ is confined to the interval $(\roprpr^2/\prmaxsup^2,\, \prmaxsup^2/\roprpr^2)$ and thus bounded away from $\roprpr^2$ or $1/\roprpr^2$ (note that restricting $\prmax$ from above is a stronger condition than restricting $\roprpr$ from above,
as can be seen from Eq.~\eqref{eq: roprpr prmax}). In fact, a simple and useful bound independent of $\RR$, comparable to that found for RR and LRR, can be obtained for OR and LOR using $\prmax$ in place of $\roprpr$.

\begin{theorem}
\label{theo: effic OR LOR, indep of RR}
The efficiency of the OR and LOR estimators is bounded for $p_1, p_2 \in (0,1)$ as
\begin{equation}
\label{eq: efic OR LOR, indep of RR}
\effic > \frac{\suc-\cerr}{\suc+\adjsuc} (1-\prmax),
\end{equation}
where $\prmax$ is defined in Eq.~\eqref{eq: prmax}. This bound is a decreasing function of $\prmax$, and
\begin{equation}
\label{eq: efic OR LOR, indep of RR, lim}
\liminf_{\prmax \rightarrow 0} \effic \geq \frac{\suc-\cerr}{\suc+\adjsuc} \geq \frac 1 {1 + \tarvar(\cerr+\adjsuc)}.
\end{equation}
\end{theorem}

\begin{figure}%
\centering%
\begin{subfigure}{\tamdosfig}%
\centering%
\includegraphics[width=\textwidth]{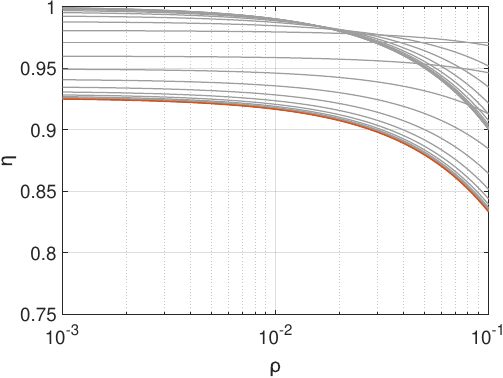}%
\caption{OR}%
\label{fig: effic OR bounds 0p04}%
\end{subfigure}%
\hfill%
\begin{subfigure}{\tamdosfig}%
\centering%
\includegraphics[width=\textwidth]{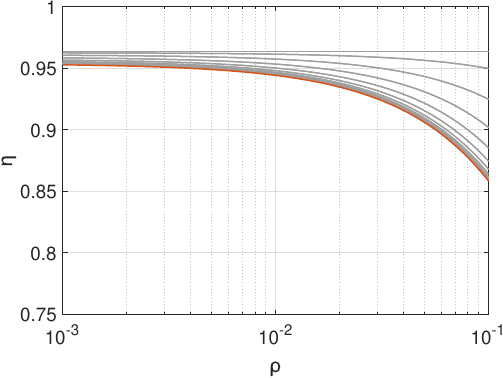}%
\caption{LOR}%
\label{fig: effic LOR bounds 0p04}%
\end{subfigure}%
\\%
\begin{subfigure}{\tamdosfig}%
\centering%
\includegraphics[scale=\legendscale]{legend_effic_bounds}%
\caption*{\vspace{-4mm}}%
\end{subfigure}%
\caption{Bounds on estimation efficiency for OR and LOR. Example for $\tarvar=0.04$}%
\label{fig: effic OR LOR bounds 0p04}%
\end{figure}%

The bound given by Theorem~\ref{theo: effic OR LOR, indep of RR}, using $\tarvar= 0.04$ as an example, is shown in Figure~\ref{fig: effic OR LOR bounds 0p04} as a function of $\prmax$ (thick, red curve). Bounds for OR and LOR obtained from Theorem~\ref{theo: effic RR LRR OR LOR} are also plotted, for the same set of values of $\RR$ as in Figure~\ref{fig: effic RR LRR bounds 0p04}, as a function of $\prmax$ (thin, grey lines), using the transformation in Eq.~\eqref{eq: roprpr prmax}. In this case, the infimum of the $\RR$-specific bounds occurs for $\RR \rightarrow 0$, irrespective of $\prmax$. The bound given by Theorem~\ref{theo: effic OR LOR, indep of RR} is equal to this infimum, as seen in the proof of that result, and as can be observed in the figure.

\begin{figure}%
\centering%
\begin{subfigure}{\tamdosfig}%
\centering%
\includegraphics[width=\textwidth]{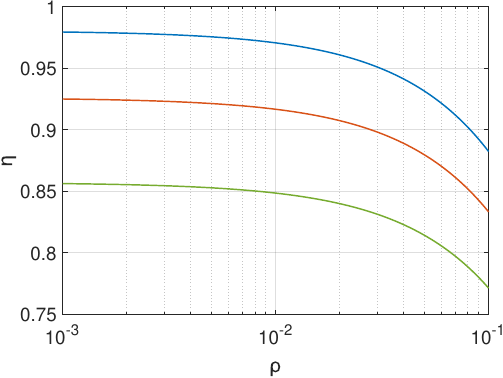}%
\caption{OR}%
\label{fig: effic OR bound}%
\end{subfigure}%
\hfill%
\begin{subfigure}{\tamdosfig}%
\centering%
\includegraphics[width=\textwidth]{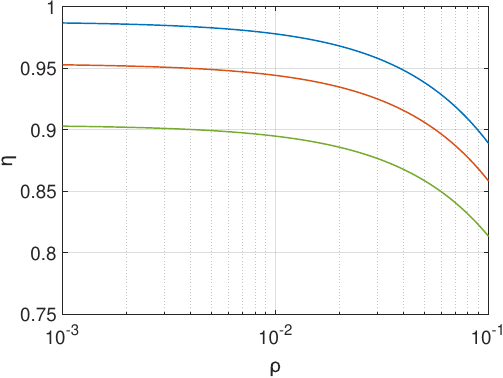}%
\caption{LOR}%
\label{fig: effic LOR bound}%
\end{subfigure}%
\\%
\begin{subfigure}{\tamdosfig}%
\centering%
\includegraphics[scale=\legendscale]{legend_effic_bound}
\caption*{\vspace{-4mm}}%
\end{subfigure}%
\caption{Bound on estimation efficiency independent of $\RR$, for OR and LOR}%
\label{fig: effic OR LOR bound}%
\end{figure}%

Figure~\ref{fig: effic OR LOR bound} shows the bound on the estimation efficiency for OR and LOR given by Theorem~\ref{theo: effic OR LOR, indep of RR}, for several values of $\tarvar$. In analogy with RR and LRR, the fact that this bound decreases with $\prmax$ means that, if $\prmax$ is assumed to be less than or equal to a given value $\prmaxsup$, the efficiency of the OR and LOR estimators will be higher than the bound particularized to $\prmaxsup$, regardless of $\RR$. For example, with $\tarvar=0.04$, if $\prmax \leq 10^{-2}$ (i.e.~if $p_1, p_2 \leq 10^{-2}$) the estimation efficiency is guaranteed to be better than $91.6\%$ and $94.4\%$ for OR and LOR respectively; and as $\prmax \rightarrow 0$ the efficiency is asymptotically higher than $92.5\%$ and $95.3\%$ respectively (same values as for RR and LRR).

In general, for all the proposed estimators, $\effic$ achieves values near $1$ for small values of $p_1$ and $p_2$, and it also increases as $\tarvar$ is reduced. Specifically, by Theorems~\ref{theo: effic RR LRR, indep of RR} and \ref{theo: effic OR LOR, indep of RR}, $\effic$ becomes better than $1/(1+\tarvar(\cerr+\alpha))$ for $\roprpr$ or $\prmax$ small enough. Thus, the estimation efficiency is high precisely \emph{when it is needed the most}, namely when the observed events are rare or when very good accuracy is desired, which is when the number of input pairs has to be large in order to guarantee the target accuracy.

\section{Conclusions}
\label{part: concl}

A procedure has been proposed to estimate the RR, LRR, OR and LOR between two populations with Bernoulli parameters $p_1$ and $p_2$. The estimators take samples in pairs, one sample from each population, in a sequential fashion. The approach consists in using these samples to generate a sequence of Bernoulli random variables with a certain parameter that is a function of $p_1$ and $p_2$, and then applying the odds or log odds estimation methods from \cite{Mendo25b} to that sequence. The resulting estimators guarantee a target accuracy, irrespective of $p_1$ and $p_2$, with accuracy understood as relative MSE for RR and OR or as MSE for LRR and LOR. The estimation efficiency, defined with respect to the \CR{} bound, is higher than that of previously proposed estimators of these parameters when $p_1$ and $p_2$
are small; and it increases when better accuracy needs to be guaranteed.

\appendix

\section{Proofs}
\label{part: proofs}

\subsection{Proof of Proposition~\ref{prop: inner RR LRR}}

The probability that Algorithm~\ref{algo: RR LRR, inner} ends at a given iteration, conditioned on it not having ended earlier, is by construction $(p_1+p_2)/2$. Therefore the number of iterations needed to produce the output, i.e.~$\vanii_1+\vanii_2$, is a geometric random variable with parameter $(p_1+p_2)/2$, and is thus finite with probability $1$.

At each iteration of the algorithm there are four possible outcomes: $i=1$, $\vai_1=0$, which happens with probability $(1-p_1)/2$; $i=2$, $\vai_2=0$, with probability $(1-p_2)/2$; $i=1$, $\vai_1=1$, with probability $p_1/2$; and $i=2$, $\vai_2=1$, with probability $p_2/2$. Conditioned on the algorithm ending at that iteration, i.e.~on the third or fourth events occurring at that iteration and not having occurred earlier, the probability that $\vao=1$ is
\[
\frac{p_1/2}{p_1/2+p_2/2} = \frac{p_1}{p_1+p_2},
\]
which equals $p$ as defined by Eq.~\eqref{eq: p RR LRR}; and the probability that $\vao=0$ is $1-p$.

To obtain $\E[\vanii_1 \cond \vao=1]$ it is convenient to first compute the probability that $\vanii_1=\nii$ and $\vao=1$, for $\nii \geq 1$. This is the probability that the inputs used by Algorithm~\ref{algo: RR LRR, inner} are as follows: $\nii-1$ tuples (finite-length sequences), with different lengths in general, such that each tuple consists of an arbitrary number (possibly $0$) of failures from $\sei_2$ followed by a failure from $\sei_1$; and then one last tuple formed by an arbitrary number (possibly $0$) of failures from $\sei_2$ and a success from $\sei_1$. The probability of the first type of tuple, considering all possible numbers of failures from $\sei_2$, is
\[
\frac{1-p_1} 2 \sum_{k=0}^{\infty} \left(\frac{1-p_2} 2\right)^k = \frac{1-p_1}{1+p_2}.
\]
Similarly, the probability of the last tuple is $p_1/(1+p_2)$. Thus, for $\nii \geq 1$,
\begin{equation}
\label{eq: Pr vanii 1 vao 1}
\Pr[\vanii_1=\nii,\, \vao=1] = \left(\frac{1-p_1}{1+p_2}\right)^{\nii-1} \frac {p_1}{1+p_2},
\end{equation}
which implies 
\begin{equation}
\label{eq: Pr vanii 1 cond vao 1}
\Pr[\vanii_1=\nii \cond \vao=1] = \frac{\Pr[\vanii_1=\nii,\, \vao=1]} {\Pr[\vao=1]} =
\left(\frac{1-p_1}{1+p_2}\right)^{\nii-1} \frac {p_1+p_2}{1+p_2}.
\end{equation}
According to Eq.~\eqref{eq: Pr vanii 1 cond vao 1}, $\vanii_1$ conditioned on $\vao=1$ has a geometric distribution with parameter $(p_1+p_2)/(1+p_2)$. This establishes Eq.~\eqref{eq: E vanii 1 cond vao 1}.

The procedure to obtain $\E[\vanii_1 \cond \vao=0]$ is analogous. In this case, the event defined by $\vanii_1=\nii$ and $\vao=0$, for $\nii \geq 0$, corresponds to Algorithm~\ref{algo: RR LRR, inner} using the following inputs: $\nii$ tuples consisting of an arbitrary number (possibly $0$) of failures from $\sei_2$ followed by a failure from $\sei_1$, and then a tuple with an arbitrary number (possibly $0$) of failures from $\sei_2$ followed by a success from $\sei_2$. This gives, for $\nii \geq 0$,
\begin{equation}
\Pr[\vanii_1=\nii \cond \vao=0] = \left(\frac{1-p_1}{1+p_2}\right)^{\nii} \frac {p_1+p_2}{1+p_2},
\end{equation}
from which Eq.~\eqref{eq: E vanii 1 cond vao 0} readily follows.

By symmetry, the arguments used in the derivation of Eq.~\eqref{eq: E vanii 1 cond vao 0} are valid if $p_1$ and $p_2$ are interchanged, $\vanii_1$ is replaced by $\vanii_2$, and the event $\vao=0$ is replaced by $\vao=1$. This implies that, for $\nii \geq 0$,
\begin{equation}
\label{eq: Pr vanii 2 cond vao 1}
\Pr[\vanii_2=\nii \cond \vao=1] = \left(\frac{1-p_2}{1+p_1}\right)^{\nii} \frac {p_1+p_2}{1+p_1},
\end{equation}
from which Eq.~\eqref{eq: E vanii 2 cond vao 1} results. Likewise, interchanging $p_1$ and $p_2$, and replacing $\vanii_1$ by $\vanii_2$ and $\vao=1$ by $\vao=0$ in Eq.~\eqref{eq: E vanii 1 cond vao 1} yields Eq.~\eqref{eq: E vanii 2 cond vao 0}.

From Eqs.~\eqref{eq: E vanii 1 cond vao 1}--\eqref{eq: E vanii 2 cond vao 0}, 
computing the value of $\E[\vanii_i]$ as $\E[\vanii_i \cond \vao=1] \Pr[\vao=1] + \E[\vanii_i \cond \vao=0] \Pr[\vao=0]$ for $i=1,2$ gives Eq.~\eqref{eq: E vanii}.

The conditional variance $\Var[\vanii_1-\vanii_2 \cond \vao=j]$, $j=0,1$ can be expressed as
\begin{equation}
\label{eq: var vanii 1 - vanii 2 cond proof}
\begin{split}
\Var[\vanii_1-\vanii_2 \cond \vao=j] &= \E[(\vanii_1-\vanii_2)^2\cond \vao=j] - \left(\E[\vanii_1\cond \vao=j] - \E[\vanii_2\cond \vao=j]\right)^2 \\
&= \E\left[\vanii_1^2\cond \vao=j\right] + \E\left[\vanii_2^2\cond \vao=j\right] - 2 \E[\vanii_1 \vanii_2 \cond \vao=j] \\
& \quad - (\E[\vanii_1\cond \vao=j] - \E[\vanii_2\cond \vao=j])^2.
\end{split}
\end{equation}
For $j=1$, the term $\E[\vanii_1^2\cond \vao=1]$ is computed from Eqs.~\eqref{eq: geom series diff diff}, \eqref{eq: E vanii 1 cond vao 1} and \eqref{eq: Pr vanii 1 cond vao 1} as
\begin{equation}
\label{eq: E vanii 1 2 cond vao 1}
\begin{split}
\E\left[\vanii_1^2\cond* \vao=1\right] &= \E[\vanii_1(\vanii_1-1)\cond \vao=1] + \E[\vanii_1\cond \vao=1] \\
&= \frac {p_1+p_2}{1+p_2} \sum_{\nii=2}^\infty \nii(\nii-1) \left(\frac{1-p_1}{1+p_2}\right)^{\nii-1} + \frac{1+p_2}{p_1+p_2} \\
&= \frac{(2-p_1+p_2)(1+p_2)}{(p_1+p_2)^2}.
\end{split}
\end{equation}
Analogously, from Eqs.~\eqref{eq: geom series diff diff}, \eqref{eq: E vanii 2 cond vao 1} and \eqref{eq: Pr vanii 2 cond vao 1},
\begin{equation}
\label{eq: E vanii 2 2 cond vao 1}
\E\left[\vanii_2^2\cond* \vao=1\right] = \frac{(2+p_1-p_2)(1-p_2)}{(p_1+p_2)^2}.
\end{equation}
Using symmetry again, $\E[\vanii_1^2\cond \vao=0]$ and $\E[\vanii_2^2\cond \vao=0]$ are obtained from Eqs.~\eqref{eq: E vanii 1 2 cond vao 1} and \eqref{eq: E vanii 2 2 cond vao 1} by exchanging $p_1$ and $p_2$, $\vanii_1$ and $\vanii_2$, as well as $\vao=1$ and $\vao=0$:
\begin{align}
\label{eq: E vanii 1 2 cond vao 0}
\E\left[\vanii_1^2\cond* \vao=0\right] = \frac{(2-p_1+p_2)(1-p_1)}{(p_1+p_2)^2}, \\
\label{eq: E vanii 2 2 cond vao 0}
\E\left[\vanii_2^2\cond* \vao=0\right] = \frac{(2+p_1-p_2)(1+p_1)}{(p_1+p_2)^2}.
\end{align}

To compute the term $\E[\vanii_1 \vanii_2\cond \vao=j]$ in Eq.~\eqref{eq: var vanii 1 - vanii 2 cond proof} it is necessary to obtain the joint distribution of $\vanii_1$ and $\vanii_2$ conditioned on $\vao$. For $j=1$, $\nii_1 \geq 1$, $\nii_2 \geq 0$, the event defined by $\vanii_1=\nii_1$, $\vanii_2=\nii_2$ and $\vao=j$ occurs when the inputs used by Algorithm~\ref{algo: RR LRR, inner} are $\nii_1-1$ failures from $\sei_1$ and $\nii_2$ failures from $\sei_2$, in any order, followed by a success from $\sei_1$. Thus
\begin{equation}
\Pr[\vanii_1=\nii_1,\, \vanii_2=\nii_2,\, \vao=1] = \binom{\nii_1+\nii_2-1}{\nii_1-1} \frac{(1-p_1)^{\nii_1-1} (1-p_2)^{\nii_2} p_1}{2^{\nii_1+\nii_2}},
\end{equation}
from which
\begin{equation}
\begin{split}
\label{eq: Pr vanii 1 vanii 2 cond vao 1}
\Pr[\vanii_1=\nii_1,\, \vanii_2=\nii_2 \cond \vao=1] &= \frac{\Pr[\vanii_1=\nii_1,\, \vanii_2=\nii_2,\, \vao=1]}{\Pr[\vao=1]} \\
&= \binom{\nii_1+\nii_2-1}{\nii_1-1} \frac{(1-p_1)^{\nii_1-1} (1-p_2)^{\nii_2} (p_1+p_2)}{2^{\nii_1+\nii_2}}.
\end{split}
\end{equation}
Similarly, for $j=0$,  $\nii_1 \geq 0$, $\nii_2 \geq 1$,
\begin{equation}
\label{eq: Pr vanii 1 vanii 2 cond vao 0}
\Pr[\vanii_1=\nii_1,\, \vanii_2=\nii_2 \cond \vao=0] = \binom{\nii_1+\nii_2-1}{\nii_1} \frac{(1-p_1)^{\nii_1} (1-p_2)^{\nii_2-1} (p_1+p_2)}{2^{\nii_1+\nii_2}}.
\end{equation}
Using Eqs.~\eqref{eq: Pr vanii 1 vanii 2 cond vao 1} and \eqref{eq: Pr vanii 1 vanii 2 cond vao 0}, and then substituting Eq.~\eqref{eq: E vanii 1 2 cond vao 0},
\begin{equation}
\label{eq: E vanii 1 vanii 2 cond vao 1}
\begin{split}
\E[\vanii_1 \vanii_2\cond \vao=1] &= \sum_{\nii_1=1}^\infty \sum_{\nii_2=1}^\infty \nii_1 \nii_2 \Pr[\vanii_1=\nii_1,\, \vanii_2=\nii_2 \cond \vao=1] \\
&= \sum_{\nii_1=1}^\infty \sum_{\nii_2=1}^\infty \nii_1 \nii_2 \binom{\nii_1+\nii_2-1}{\nii_1-1} \frac{(1-p_1)^{\nii_1-1} (1-p_2)^{\nii_2} (p_1+p_2)}{2^{\nii_1+\nii_2}} \\
&= \sum_{\nii_1=1}^\infty \sum_{\nii_2=1}^\infty \nii_1^2 \binom{\nii_1+\nii_2-1}{\nii_1} \frac{(1-p_1)^{\nii_1-1} (1-p_2)^{\nii_2} (p_1+p_2)}{2^{\nii_1+\nii_2}} \\
&= \frac{1-p_2}{1-p_1} \E\left[\vanii_1^2 \cond* \vao=0\right] = \frac{(2-p_1+p_2)(1-p_2)}{(p_1+p_2)^2}.
\end{split}
\end{equation}
By an analogous argument,
\begin{equation}
\label{eq: E vanii 1 vanii 2 cond vao 0}
\E[\vanii_1 \vanii_2\cond \vao=0] = \frac{1-p_1}{1-p_2} \E\left[\vanii_2^2 \cond* \vao=1\right] = \frac{(2+p_1-p_2)(1-p_1)}{(p_1+p_2)^2}.
\end{equation}

Substituting Eqs.~\eqref{eq: E vanii 1 cond vao 1}, \eqref{eq: E vanii 2 cond vao 1}, \eqref{eq: E vanii 1 2 cond vao 1}, \eqref{eq: E vanii 2 2 cond vao 1} and \eqref{eq: E vanii 1 vanii 2 cond vao 1} into Eq.~\eqref{eq: var vanii 1 - vanii 2 cond proof} for $j=1$, the right-hand side of Eq.~\eqref{eq: var vanii 1 - vanii 2 cond} is obtained. The expression for $j=0$ is the same, as can be seen substituting Eq.~\eqref{eq: E vanii 1 cond vao 0}, \eqref{eq: E vanii 2 cond vao 0}, \eqref{eq: E vanii 1 2 cond vao 0}, \eqref{eq: E vanii 2 2 cond vao 0} and \eqref{eq: E vanii 1 vanii 2 cond vao 0} into Eq.~\eqref{eq: var vanii 1 - vanii 2 cond proof}, or simply noting that $\Var[\vanii_1-\vanii_2 \cond \vao=j]$ is symmetric to an exchange of $\vanii_1$ and $\vanii_2$ and the right-hand side of Eq.~\eqref{eq: var vanii 1 - vanii 2 cond} is symmetric to an exchange of $p_1$ and $p_2$.
\qed

\subsection{Proof of Proposition~\ref{prop: inner OR LOR}}

The numbers of input samples $\vanii_1$ and $\vanii_2$ coincide with the number of iterations of Algorithm~\ref{algo: OR LOR, inner}, which is, by construction, a geometric random variable with parameter $p_1(1-p_2)+p_2(1-p_1)$. This implies that $\vanii_i$, $i=1,2$ is finite with probability $1$, and that $\E[\vanii_i]$ is given by the right-hand side of Eq.~\eqref{eq: E vanii OR LOR}.

The probability that the algorithm ends at the $\nii$-th iteration and outputs $\vao=1$ is
\begin{equation}
\label{eq: Pr vanii vao 1 OR LOR}
\Pr[\vanii_1 = \nii,\, \vao=1] = (1-p_1(1-p_2)-p_2(1-p_1))^{\nii-1} p_1(1-p_2),
\end{equation}
and similarly, the probability that it ends at the $\nii$-th iteration and outputs $\vao=0$ is
\begin{equation}
\label{eq: Pr vanii vao 0 OR LOR}
\Pr[\vanii_1 = \nii,\, \vao=0] = (1-p_1(1-p_2)-p_2(1-p_1))^{\nii-1} p_2(1-p_1).
\end{equation}
Therefore,
\begin{equation}
\label{Pr vao=1 cond div vao=0 cond}
\frac{\Pr[\vao=1 \cond \vanii_1 = \nii]}{\Pr[\vao=0 \cond \vanii_1 = \nii]} = \frac{\Pr[\vanii_1 = \nii,\, \vao=1]}{\Pr[\vanii_1 = \nii,\, \vao=0]} = \frac{p_1(1-p_2)}{p_2(1-p_1)}.
\end{equation}
Since the result of Eq.~\eqref{Pr vao=1 cond div vao=0 cond} is independent of $\nii$, it follows that $\Pr[\vao=1] / \Pr[\vao=0]$ equals $p_1(1-p_2)/(p_2(1-p_1))$, from which
\begin{equation}
\label{eq: Pr vao 1 p OR LOR}
\Pr[\vao=1] = 1-\Pr[\vao=0] = p,
\end{equation}
with $p$ given by Eq.~\eqref{eq: p OR LOR}.

Using Eqs.~\eqref{eq: Pr vanii vao 1 OR LOR}, \eqref{eq: Pr vanii vao 0 OR LOR} and \eqref{eq: Pr vao 1 p OR LOR},
\begin{equation}
\frac{\Pr[\vanii_1 = \nii \cond \vao=1]}{\Pr[\vanii_1 = \nii \cond \vao=0]} = 
\frac{\Pr[\vanii_1 = \nii,\, \vao=1]\, \Pr[\vao=0]}{\Pr[\vanii_1 = \nii,\, \vao=0]\, \Pr[\vao=1]} =
\frac{ p_1(1-p_2) (1-p)}{p_2(1-p_1) p} = 1.
\end{equation}
 This implies that $\E[\vanii_i \cond \vao=1] = \E[\vanii_i \cond \vao=0] = \E[\vanii_i]$, $i=1,2$, which completes the proof of Eq.~\eqref{eq: E vanii OR LOR}.
\qed

\subsection{Proof of Theorem~\ref{theo: MSE}}

The results follow from Propositions~\ref{prop: inner RR LRR} and \ref{prop: inner OR LOR} and from  \cite[theorems~1 and~3]{Mendo25b}.
\qed

\subsection{Proof of Proposition~\ref{prop: outer RR LRR}}

To obtain the joint probability function of $\vanoi_1$, $\vanoi_2$, it is convenient to first compute that of $\vanoi_1$, $\vanoi_2$, $\vanoo'$, $\vanoo''$, where $\vanoo'$ and $\vanoo''$ are the numbers of samples from $\seo$ used by the two IBS procedures in Algorithm~\ref{algo: outer}.

There are two limitations on the values that the above variables can have. First, the numbers of samples from $\seo$ used by the two IBS processes, i.e.~$\vanoo'$ and $\vanoo''$, are at least $\suc+\adjsuc$ and $\suc-\adjsuc$ respectively. Second, the number of observations taken from $\sei_1$, i.e.~$\vanoi_1$, is necessarily greater than or equal to the total number of successes from $\seo$ used by the two IBS processes, which is $\suc+\adjsuc+\vanoo''- (\suc-\adjsuc) = \vanoo''+2\adjsuc$; and similarly $\vanoi_2$ must be greater than or equal to $\vanoo'-2\adjsuc$. Thus, $\Pr[\vanoi_1 = \noi_1, \vanoi_2 = \noi_2, \vanoo'=\noo', \vanoo''=\noo'']$ will only be non-zero if
\begin{align}
\label{eq: noo', restr}
\suc+\adjsuc &\leq \noo' \leq \noi_2+2\adjsuc, \\
\label{eq: noo'', restr}
\suc-\adjsuc &\leq \noo'' \leq \noi_1-2\adjsuc.
\end{align}

Consider $\noi_1$, $\noi_2$, $\noo'$ and $\noo''$ that satisfy the above restrictions. In accordance with these values, in step \ref{step: outer: IBS 1} of Algorithm~\ref{algo: outer}, $\noo'$ samples of $\seo$ are generated, from which $\suc+\adjsuc$ are successes and $\noo'-\suc-\adjsuc$ are failures; and in step \ref{step: outer: IBS 2}, $\noo''$ samples of $\seo$ are generated, from which $\suc-\adjsuc$ are failures and $\noo''-\suc+\adjsuc$ are successes. These samples of $\seo$ are generated using Algorithm~\ref{algo: RR LRR, inner}, which requires $\noi_1$ observations from $\sei_1$ and $\noi_2$ observations from $\sei_2$ in total. Of the $\noi_1$ observations from $\sei_1$, $\noo''+2\adjsuc$ are successes and $\noi_1-\noo''-2\adjsuc$ are failures; and similarly, of the $\noi_2$ observations from $\sei_2$, $\noo'-2\adjsuc$ are successes and $\noi_2-\noo'+2\adjsuc$ are failures.

It is convenient, for the moment, to view each observation taken as input by Algorithm~\ref{algo: outer} as belonging to one of three categories: failures from $\sei_1$, failures from $\sei_2$, or successes from either sequence. The last observation is necessarily a success (specifically a success from $\sei_2$, because it ends the second IBS process in the outer loop of Algorithm~\ref{algo: outer}); and the preceding observations are $\noi_1-\noo''-2\adjsuc$ failures from $\sei_1$, $\noi_2-\noo'+2\adjsuc$ failures from $\sei_2$ and $\noo'+ \noo''-1$ successes, all of which can be arranged in any order. There are thus
\[
\binom{\noi_1+\noi_2-1}{\noi_1-\noo''-2\adjsuc,\, \noi_2-\noo'+2\adjsuc,\, \noo'+\noo''-1}
\]
possible arrangements (distinct permutations of the three categories).

The third category defined above can at this point be split into two, namely successes from $\sei_1$ or from $\sei_2$, as follows. Given an arrangement of the $\noo'+ \noo''-1$ successes within the total of $\noi_1+\noi_2-1$ input observations, there are a number of possible internal orders of successes from $\sei_1$ and from $\sei_2$. This corresponds to the order of the successes and failures from $\seo$ used by the two IBS procedures. Namely, the first IBS process consumes $\noo'$ samples from $\seo$, of which $\suc+\adjsuc$ are successes. The last one is a success, and the rest can be arranged in any order, which gives
\[
\binom{\noo'-1}{\suc+\adjsuc-1}
\]
possibilities. Similarly, the second IBS process consumes $\noo''$ samples from $\seo$, of which $\suc-\adjsuc$ are failures. The last one is necessarily a failure (in accordance with the last observed input being a success of $\sei_2$), and the rest can be arranged in
\[
\binom{\noo''-1}{\suc-\adjsuc-1}
\]
possible ways.

Based on the above, considering the four categories defined by successes or failures of either $\sei_1$ or $\sei_2$, the number of allowed arrangements of these four categories for $\vanoi_1 = \noi_1$, $\vanoi_2 = \noi_2$, $\vanoo'=\noo'$ and $\vanoo''=\noo''$ is
\[
\binom{\noi_1+\noi_2-1}{\noi_1-\noo''-2\adjsuc,\, \noi_2-\noo'+2\adjsuc,\, \noo'+\noo''-1} \binom{\noo'-1}{\suc+\adjsuc-1} \binom{\noo''-1}{\suc-\adjsuc-1}.
\]
Each such arrangement contains $\noo''+2\adjsuc$ successes from $\sei_1$, $\noo'-2\adjsuc$ successes from $\sei_2$, $\noi_1-\noo''-2\adjsuc$ failures from $\sei_1$ and $\noi_2-\noo'+2\adjsuc$ failures from $\sei_2$. According to Algorithm~\ref{algo: RR LRR, inner}, the probabilities of the input observation being a success from $\sei_1$, a success from $\sei_2$, a failure from $\sei_1$ or a failure from $\sei_2$ are respectively $p_1/2$, $p_2/2$, $(1-p_1)/2$ and $(1-p_2)/2$. Therefore the joint probability function of $\vanoi_1$, $\vanoi_2$, $\vanoo'$, $\vanoo''$ is
\begin{equation}
\label{eq: Pr vanoi 1 vanoi 2 vanoo' vanoo'' RR LRR}
\begin{split}
& \Pr\left[\vanoi_1 = \noi_1, \vanoi_2 = \noi_2, \vanoo'=\noo', \vanoo''=\noo''\right] \\
& \quad = \binom{\noi_1+\noi_2-1}{\noi_1-\noo''-2\adjsuc,\, \noi_2-\noo'+2\adjsuc,\, \noo'+\noo''-1} \binom{\noo'-1}{\suc+\adjsuc-1} \binom{\noo''-1}{\suc-\adjsuc-1} \\
& \quad\quad \cdot \frac{(1-p_1)^{\noi_1-\noo''-2\adjsuc} (1-p_2)^{\noi_2-\noo'+2\adjsuc} p_1^{\noo''+2\adjsuc} p_2^{\noo'-2\adjsuc}}{2^{\noi_1+\noi_2}}
\end{split}
\end{equation}
when $\noi_1$, $\noi_2$, $\noo'$, $\noo''$ satisfy the restrictions given by Eqs.~\eqref{eq: noo', restr} and \eqref{eq: noo'', restr}, 
and otherwise it equals $0$. In consequence,
\begin{equation}
\Pr\left[\vanoi_1 = \noi_1, \vanoi_2 = \noi_2\right] = 
\sum_{\noo'=\suc+\adjsuc}^{\noi_2+2\adjsuc} \, \sum_{\noo''=\suc-\adjsuc}^{\noi_1-2\adjsuc} \Pr\left[\vanoi_1 = \noi_1, \vanoi_2 = \noi_2, \vanoo'=\noo', \vanoo''=\noo''\right]
\end{equation}
for $\noi_1 \geq \suc+\adjsuc$, $\noi_2 \geq \suc-\adjsuc$, which combined with Eq.~\eqref{eq: Pr vanoi 1 vanoi 2 vanoo' vanoo'' RR LRR} gives Eq.~\eqref{eq: Pr vanoi 1 vanoi 2 RR LRR}.

Using Proposition~\ref{prop: inner RR LRR}, the mean of $\vanoi_1$ conditioned on $\vanoo'$, $\vanoo''$ is obtained as
\begin{equation}
\begin{split}
\label{eq: vanoi 1 cond vanoo' vanoo''}
\E\left[\vanoi_1 \cond* \vanoo', \vanoo''\right] &= \left(\vanoo''+2\adjsuc\right) \E[\vanii_1 \cond \vao=1] + \left(\vanoo'-2\adjsuc\right) \E[\vanii_1 \cond \vao=0] \\
&= \left(\vanoo''+2\adjsuc\right) \frac{1+p_2}{p_1+p_2} + \left(\vanoo'-2\adjsuc\right) \frac{1-p_1}{p_1+p_2}.
\end{split}
\end{equation}
From Eqs.~\eqref{eq: E N neg bin} and \eqref{eq: p RR LRR} it stems that
$\E[\vanoo'] = (\suc+\adjsuc)(p_1+p_2)/p_1$ and $\E[\vanoo''] = (\suc-\adjsuc)(p_1+p_2)/p_2$, which combined with Eq.~\eqref{eq: vanoi 1 cond vanoo' vanoo''} give $\E[\vanoi_1]$ as in Eq.~\eqref{eq: vanoi 1 RR LRR}. By an analogous argument, the same expression is obtained for $\E[\vanoi_2]$.
\qed

\subsection{Proof of Proposition~\ref{prop: E max cota RR LRR}}

Using Eqs.~\eqref{eq: vanoi max} and \eqref{eq: max suma dif abs}, $\E[\vanoi]$ can be written as
\begin{equation}
\label{eq: E max suma dif abs}
\E[\vanoi] = \frac{\E[\vanoi_1 + \vanoi_2]} 2 + \frac{\E[|\vanoi_1 - \vanoi_2|]} 2.
\end{equation}
From Eq.~\eqref{eq: vanoi 1 RR LRR} in Proposition~\ref{prop: outer RR LRR} it stems that $\E[\vanoi_1-\vanoi_2] = 0$, and then using Jensen's inequality \cite[theorem~7.5]{Lehmann98} it is easy to see that $\E[|\vanoi_1-\vanoi_2|] < \sqrt{\Var[\vanoi_1-\vanoi_2]}$, which substituted into Eq.~\eqref{eq: E max suma dif abs} gives
\begin{equation}
\label{eq: vanoi <}
\E[\vanoi] < \frac{\E[\vanoi_1 + \vanoi_2]} 2 + \frac{\sqrt{\Var[\vanoi_1-\vanoi_2]}} 2.
\end{equation}
The term $\E[\vanoi_1 + \vanoi_2]$ is obtained using Eq.~\eqref{eq: vanoi 1 RR LRR} again:
\begin{equation}
\label{eq: E vanoi 1 + vanoi 2}
\E[\vanoi_1 + \vanoi_2] = 2\left(\frac{\suc+\adjsuc}{p_1} + \frac{\suc-\adjsuc}{p_2}\right).
\end{equation}
To compute $\Var[\vanoi_1-\vanoi_2]$, it is helpful to condition on the numbers of samples of $\seo$ used by the two IBS procedures, i.e.~$\vanoo'$, $\vanoo''$, and apply the law of total variance \cite[theorem~12.2.6]{Athreya06}:
\begin{equation}
\label{eq: Var vanoi 1 - vanoi 2}
\Var[\vanoi_1-\vanoi_2] = \E\left[\Var\left[\vanoi_1-\vanoi_2 \cond* \vanoo', \vanoo''\right]\right] + \Var\left[\E\left[\vanoi_1-\vanoi_2 \cond* \vanoo', \vanoo''\right]\right].
\end{equation}

The first IBS procedure in Algorithm~\ref{algo: outer} uses $\vanoo'$ samples from $\seo$, of which $\suc+\adjsuc$ are successes and $\vanoo'-\suc-\adjsuc$ are failures. Similarly, the second IBS procedure uses $\vanoo''$ samples from $\seo$, of which $\vanoo''-\suc+\adjsuc$ are successes and $\suc-\adjsuc$ are failures. Thus, the estimator uses $\vanoo''+2\adjsuc$ successes and $\vanoo'-2\adjsuc$ failures from $\seo$ in total. Each sample from $\seo$ is generated by Algorithm~\ref{algo: RR LRR, inner}. Since different executions of the algorithm are independent,
\begin{multline}
\label{eq: var vanoi 1 - vanoi 2 cond}
\Var\left[\vanoi_1-\vanoi_2 \cond* \vanoo', \vanoo''\right] \\= \left(\vanoo''+2\adjsuc\right) \Var[\vanii_1-\vanii_2 \cond \vao=1] + \left(\vanoo'-2\adjsuc\right) \Var[\vanii_1-\vanii_2 \cond \vao=0],
\end{multline}
where $\vanii_1$, $\vanii_2$ are the numbers of inputs used by a single run of Algorithm~\ref{algo: RR LRR, inner}. Substituting Eq.~\eqref{eq: var vanii 1 - vanii 2 cond} from Proposition~\ref{prop: inner RR LRR} into Eq.~\eqref{eq: var vanoi 1 - vanoi 2 cond} yields
\begin{equation}
\label{eq: var vanoi 1 - vanoi 2 cond bis}
\Var\left[\vanoi_1-\vanoi_2 \cond* \vanoo', \vanoo''\right] = \frac{2 \left(\vanoo'+\vanoo''\right)(p_1+p_2 - 2 p_1 p_2)}{(p_1+p_2)^2}.
\end{equation}
Therefore, computing $\E[\vanoo']$ and $\E[\vanoo'']$ from Eqs.~\eqref{eq: E N neg bin} and \eqref{eq: p RR LRR},
\begin{equation}
\begin{split}
\label{eq: E var vanoi 1 - vanoi 2 cond}
\E\left[\Var\left[\vanoi_1-\vanoi_2 \cond* \vanoo', \vanoo''\right]\right] &= \frac{2(p_1+p_2 - 2 p_1 p_2)} {p_1+p_2} \left( \frac{\suc+\adjsuc}{p_1} + \frac{\suc-\adjsuc}{p_2} \right) \\
&= 2 \left( \frac{\suc+\adjsuc}{p_1} + \frac{\suc-\adjsuc}{p_2} \right) - 4 \suc + \frac{4\adjsuc(p_1-p_2)}{p_1+p_2}.
\end{split}
\end{equation}
As for the second summand in Eq.~\eqref{eq: Var vanoi 1 - vanoi 2}, $\E[\vanoi_1-\vanoi_2 \cond \vanoo', \vanoo'']$ can be obtained as
\begin{equation}
\label{eq: E vanoi 1 - vanoi 2 cond}
\begin{split}
\E\left[\vanoi_1-\vanoi_2 \cond* \vanoo', \vanoo''\right] &= \left(\vanoo''+2\adjsuc\right) \E[\vanii_1-\vanii_2 \cond \vao=1] \\
&\quad + \left(\vanoo'-2\adjsuc\right) \E[\vanii_1-\vanii_2 \cond \vao=0].
\end{split}
\end{equation}
Making use of Proposition~\ref{prop: inner RR LRR} again, Eq.~\eqref{eq: E vanoi 1 - vanoi 2 cond} becomes
\begin{equation}
\label{eq: E vanoi 1 - vanoi 2 cond bis}
\E\left[\vanoi_1-\vanoi_2 \cond* \vanoo', \vanoo''\right] = \frac{2\left(\vanoo''p_2-\vanoo'p_1\right)} {p_1+p_2}  + 4\adjsuc,
\end{equation}
and then, computing  $\Var[\vanoo']$ and $\Var[\vanoo'']$ from Eqs.~\eqref{eq: E Var neg bin} and \eqref{eq: p RR LRR},
\begin{equation}
\label{eq: Var E vanoi 1 - vanoi 2 cond}
\Var\left[\E\left[\vanoi_1-\vanoi_2 \cond* \vanoo', \vanoo''\right]\right] = \frac{4((\suc-\adjsuc)p_1 + (\suc+\adjsuc)p_2)} {p_1+p_2} = 4 \suc - \frac{4\adjsuc(p_1-p_2)}{p_1+p_2}.
\end{equation}
According to Eqs.~\eqref{eq: Var vanoi 1 - vanoi 2}, \eqref{eq: E var vanoi 1 - vanoi 2 cond} and \eqref{eq: Var E vanoi 1 - vanoi 2 cond},
\begin{equation}
\label{eq: Var vanoi 1 - vanoi 2 bis}
\Var\left[\vanoi_1-\vanoi_2 \right] = 2 \left( \frac{\suc+\adjsuc}{p_1} + \frac{\suc-\adjsuc}{p_2} \right).
\end{equation}
Combining Eqs.~\eqref{eq: vanoi <}, \eqref{eq: E vanoi 1 + vanoi 2} and \eqref{eq: Var vanoi 1 - vanoi 2 bis} yields Eq.~\eqref{eq: E vanoi < RR LRR}.

Substituting Eqs.~\eqref{eq: vanoi 1 RR LRR} and \eqref{eq: E vanoi < RR LRR} into Eq.~\eqref{eq: sef} and taking into account Eqs.~\eqref{eq: roprpr} and \eqref{eq: RR} gives Eq.~\eqref{eq: sef >}. Lastly, Eq.~\eqref{eq: lim sef} follows from Eq.~\eqref{eq: sef >} and the fact that $\sef \leq 1$.
\qed

\subsection{Proof of Proposition~\ref{prop: outer OR LOR}}

Using Proposition~\ref{prop: inner OR LOR}, $\E[\vanoi_i]$, $i=1,2$ can be computed as
\begin{equation}
\label{eq: E vanoi OR LOR proof}
\E[\vanoi_i] = \E\left[\vanoo'+\vanoo''\right] \E\left[\vanii_i\right] = \frac { \E\left[\vanoo'\right]+\E\left[\vanoo''\right]} {p_1(1-p_2) + p_2(1-p_1)}.
\end{equation}
From Eq.~\eqref{eq: E N neg bin}, the terms $\E[\vanoo']$ and $\E[\vanoo'']$ equal $(\suc+\adjsuc)/p$ and $(\suc-\adjsuc)/(1-p)$ respectively, with $p$ defined by Eq.~\eqref{eq: p OR LOR}. Substituting this into Eq.~\eqref{eq: E vanoi OR LOR proof} yields Eq.~\eqref{eq: vanoi OR LOR}.
\qed

\subsection{Proof of Theorem~\ref{theo: effic RR LRR OR LOR}}

The RR estimator is considered first. Particularizing Eq.~\eqref{eq: efic gen} for RR, that is $\partial \parg/\partial p_1 = \parg/p_1$, $\partial \parg/\partial p_2 = -\parg/p_2$, gives
\begin{equation}
\label{eq: efic RR}
\effic = \frac{\displaystyle \left( \frac 1 {p_1} + \frac 1 {p_2} - 2 \right) \parg^2} {\E[\vanoi] \Var[\vahparg]}.
\end{equation}
Using Eqs.~\eqref{eq: sef}, \eqref{eq: roprpr} and \eqref{eq: RR}, as well as Eq.~\eqref{eq: vanoi 1 RR LRR} from Proposition~\ref{prop: outer RR LRR}, this becomes
\begin{equation}
\label{eq: efic RR bis}
\effic
= \frac{(p_1 + p_2 - 2 p_1 p_2) \parg^2 \sef } {\left( (\suc+\adjsuc)p_2 + (\suc-\adjsuc)p_1 \right) \Var[\vahparg]}
= \frac{\left(\displaystyle \frac 1 {\sqrt{\RR}} + \sqrt{\RR}-2\roprpr \right) \parg^2 \sef } {\left( \displaystyle \frac{\suc+\adjsuc}{\sqrt{\RR}} + (\suc-\adjsuc)\sqrt{\RR} \right) \Var[\vahparg]}.
\end{equation}
Combining Eq.~\eqref{eq: efic RR bis} with Eq.~\eqref{eq: rel MSE < RR OR} from Theorem~\ref{theo: MSE}, and taking into account Eq.~\eqref{eq: p RR LRR} and the definitions in Eqs.~\eqref{eq: cerr} and \eqref{eq: vef} for RR, the estimator of this parameter is seen to satisfy Eq.~\eqref{eq: efic RR LRR}.

The proof for LRR is analogous. In this case $\partial \parg/\partial p_1 = 1/p_1$, $\partial \parg/\partial p_2 = -1/p_2$, and Eq.~\eqref{eq: MSE < LRR LOR} from Theorem~\ref{theo: MSE} is used. The same bound for $\effic$ is obtained, only with $\cerr$ and $\vef$ defined differently, according to Eqs.~\eqref{eq: cerr} and \eqref{eq: vef}.

For OR, since $\partial \parg/\partial p_1 = \parg/(p_1(1-p_1))$, $\partial \parg/\partial p_2 = -\parg/(p_2(1-p_2))$, Eq.~\eqref{eq: efic gen} gives
\begin{equation}
\label{eq: efic OR}
\effic = \frac{\left( \displaystyle \frac 1 {p_1(1-p_1)} + \frac 1 {p_2(1-p_2)} \right) \parg^2} {\E[\vanoi] \Var[\vahparg]}.
\end{equation}
Noting that $\sef=1$ in this case, and using Eqs.~\eqref{eq: roprpr} and \eqref{eq: RR}, as well as Proposition~\ref{prop: outer OR LOR},
\begin{equation}
\label{eq: efic OR bis}
\begin{split}
\effic & = \frac{\left(p_1(1-p_1) + p_2(1-p_2)\right) \parg^2} {\left((\suc+\adjsuc)p_2(1-p_1) + (\suc-\adjsuc)p_1(1-p_2)\right) \Var[\vahparg]} \\
&= \frac{\left( \displaystyle \frac 1 {\sqrt{\RR}} + \sqrt{\RR} - \roprpr \left(\displaystyle \frac 1 {\RR} + \RR \right) \right) \parg^2} {\left( (\suc+\adjsuc)\left(\displaystyle \frac 1 {\sqrt{\RR}} - \roprpr\right) + (\suc-\adjsuc)\left(\sqrt{\RR}-\roprpr\right) \right) \Var[\vahparg]},
\end{split}
\end{equation}
which using Eq.~\eqref{eq: MSE < LRR LOR} from Theorem~\ref{theo: MSE}, as well as Eq.~\eqref{eq: p OR LOR} and the definitions in Eqs.~\eqref{eq: cerr} and \eqref{eq: vef} for OR, yields Eq.~\eqref{eq: efic OR LOR} for the estimator of this parameter.

The proof for LOR is analogous to that for OR. The same bound for $\effic$ is obtained as in that case, with the corresponding definitions for $\cerr$ and $\vef$.
\qed

\subsection{Proof of Proposition~\ref{prop: RR roprpr prmaxsup}}

The first part is immediate from Eq.~\eqref{eq: roprpr prmax}. As for the second part, $\prmax < \prmaxsup$ implies that $p_1, p_2 < \prmaxsup$, and Eqs.~\eqref{eq: roprpr} and \eqref{eq: RR} give $p_1 = \roprpr \sqrt{\RR}$ and $p_2 = \roprpr/\sqrt{\RR}$, from which the result follows.
\qed

\subsection{Proof of Theorem~\ref{theo: effic RR LRR, indep of RR}}

From Proposition~\ref{prop: RR roprpr prmaxsup} with $\prmaxsup = 1$, for a given $\roprpr$ the possible values of $\RR$ are restricted to the interval $(\roprpr^2, 1/\roprpr^2)$. Then, for RR and LRR, defining
\begin{align}
\label{eq: efficm RR LRR}
\efficm &= \frac
{\displaystyle\frac 1 {\sqrt{\RR}}+\sqrt{\RR}-2\roprpr}
{\displaystyle\frac{\suc+\adjsuc}{\sqrt{\RR}} + (\suc-\adjsuc)\sqrt{\RR}}, \\
\label{eq: sefm RR LRR}
\sefm &= \frac 1 {1 + \sqrt{ \displaystyle \frac {\roprpr} {2\left( (\suc+\adjsuc)/\sqrt{\RR} + (\suc-\adjsuc)\sqrt{\RR} \right)} } },
\end{align}
and using the fact that $\vef<1$, it follows from Eq.~\eqref{eq: efic RR LRR} in Theorem~\ref{theo: effic RR LRR OR LOR} and from Eq.~\eqref{eq: sef >} in Proposition~\ref{prop: E max cota RR LRR} that
\begin{equation}
\label{eq: effic > inf inf}
\effic > (\suc-\cerr) \inf_{\RR \in (\roprpr^2, 1/\roprpr^2)} \efficm \: \inf_{\RR \in (\roprpr^2, 1/\roprpr^2)} \sefm.
\end{equation}
Differentiating $\efficm$ with respect to $\sqrt{\RR}$ gives
\begin{equation}
\label{eq: partial efficm partial sqrt RR}
\frac{\partial \efficm}{\partial \sqrt{\RR}} = \frac {2\roprpr(\suc-\adjsuc)\RR+4\adjsuc\sqrt{\RR}-2\roprpr(\suc+\adjsuc)} {(\suc+\adjsuc+(\suc-\adjsuc)\RR)^2}.
\end{equation}
Using Eq.~\eqref{eq: partial efficm partial sqrt RR}, and taking into account that $\sqrt{\RR}$ is positive, it is easily seen that $\efficm$ has a single minimum at
\begin{equation}
\label{eq: sqrt RR arginf}
\sqrt{\RR} = \frac{-\adjsuc + \sqrt{\adjsuc^2 + \roprpr^2(\suc^2-\adjsuc^2)}}{\roprpr(\suc-\adjsuc)},
\end{equation}
which corresponds to
\begin{equation}
\label{eq: 1 / sqrt RR arginf}
\frac 1 {\sqrt{\RR}} = \frac{\adjsuc + \sqrt{\adjsuc^2 + \roprpr^2(\suc^2-\adjsuc^2)}}{\roprpr(\suc+\adjsuc)}.
\end{equation}
From Eqs.~\eqref{eq: efficm RR LRR}, \eqref{eq: sqrt RR arginf} and \eqref{eq: 1 / sqrt RR arginf} it follows that
\begin{equation}
\label{eq: efficm RR LRR 2}
\inf_{\RR \in (\roprpr^2, 1/\roprpr^2)} \efficm \geq \frac{\suc-\sqrt{\adjsuc^2+\roprpr^2(\suc^2-\adjsuc^2)}}{\suc^2-\adjsuc^2}.
\end{equation}
Similarly, differentiating $\sefm$ with respect to $\sqrt{\RR}$, it can be seen that it has a single minimum at
\begin{equation}
\label{eq: RR arginf}
\RR = \frac{\suc+\adjsuc}{\suc-\adjsuc},
\end{equation}
and from Eqs.~\eqref{eq: sefm RR LRR} and \eqref{eq: RR arginf} it follows that
\begin{equation}
\label{eq: secm RR LRR 2}
\inf_{\RR \in (\roprpr^2, 1/\roprpr^2)} \sefm \geq \frac{1}{1 + \displaystyle \frac 1 2 \sqrt{ \frac{\roprpr}{\sqrt{\suc^2-\adjsuc^2}}}} = \frac {2\sqrt[4]{\suc^2-\adjsuc^2}} {2\sqrt[4]{\suc^2-\adjsuc^2} + \sqrt{\roprpr}}.
\end{equation}
Combining Eqs.~\eqref{eq: effic > inf inf}, \eqref{eq: efficm RR LRR 2} and \eqref{eq: secm RR LRR 2} gives Eq.~\eqref{eq: efic RR LRR, indep of RR}.

The right-hand side of Eq.~\eqref{eq: efic RR LRR, indep of RR} decreases with $\roprpr$, and tends to $(\suc-\cerr)/(\suc+\adjsuc)$ as $\roprpr \rightarrow 0$. This establishes the first inequality in Eq.~\eqref{eq: efic RR LRR, indep of RR, lim}, and then the second is obtained using Eq.~\eqref{eq: suc tarvar}.
\qed

\subsection{Proof of Theorem~\ref{theo: effic OR LOR, indep of RR}}

In view of Eq.~\eqref{eq: efic OR LOR} from Theorem~\ref{theo: effic RR LRR OR LOR}, let
\begin{equation}
\label{eq: efficm OR LOR}
\efficm = \frac{\displaystyle \frac 1 {\sqrt{\RR}} + \sqrt{\RR} - \roprpr \left(\displaystyle \frac 1 {\RR} + \RR \right)} {(\suc+\adjsuc)\left(\displaystyle \frac 1 {\sqrt{\RR}} - \roprpr\right) + (\suc-\adjsuc)\left(\sqrt{\RR}-\roprpr\right)}.
\end{equation}
Then, taking into account that $\vef<1$, to establish Eq.~\eqref{eq: efic OR LOR, indep of RR} it suffices to show that
\begin{equation}
\label{eq: efficm get OR LOR}
\efficm \geq \frac{1-\prmax}{\suc+\adjsuc}.
\end{equation}

Assume $\RR \leq 1$. In these conditions, Eq.~\eqref{eq: roprpr prmax} reduces to $\roprpr= \prmax \sqrt{\RR}$, and Eq.~\eqref{eq: efficm OR LOR} becomes
\begin{equation}
\label{eq: efficm OR LOR 2}
\efficm = \frac {1-\prmax+\RR-\prmax\RR^2} {(\suc+\adjsuc)(1-\prmax\RR) + (\suc-\adjsuc)(1-\prmax)\RR}
= \frac {1-\prmax+\RR-\prmax\RR^2} {((1-2\prmax)\suc-\adjsuc)\RR + \suc+\adjsuc}.
\end{equation}
Differentiating with respect to $\RR$,
\begin{equation}
\label{eq: der efficm OR LOR}
\frac{\partial \efficm}{\partial \RR} = \frac
{-\prmax((1-2\prmax)\suc-\adjsuc)\RR^2 - 2\prmax(\suc+\adjsuc)\RR + \prmax(3-2\prmax)\suc+(2-\prmax)\adjsuc}
{\left(((1-2\prmax)\suc-\adjsuc)\RR + \suc+\adjsuc\right)^2}.
\end{equation}
It will be useful in the following to note that
\begin{equation}
\label{eq: der efficm OR LOR at 1}
\left. \frac{\partial \efficm}{\partial \RR} \right\rfloor_{\RR=1} = \frac{\adjsuc}{2\suc^2(1-\prmax)} \geq 0.
\end{equation}
The coefficient of $\RR^2$ in the numerator of Eq.~\eqref{eq: der efficm OR LOR} is positive, negative or zero depending on whether $\prmax$ is greater, smaller or equal to $(\suc-\adjsuc)/(2\suc)$ respectively. The coefficient of $\RR$ is always negative, and the independent term is always positive.

According to the above, three cases need to be distinguished. For $\prmax > (\suc-\adjsuc)/(2\suc)$ the numerator of Eq.~\eqref{eq: der efficm OR LOR} is an upward-opening parabola. This parabola has two positive roots, according to Descartes' rule of signs \cite{Komornik06}; and its minimum occurs at $\RR = (\suc+\adjsuc) / ((2\prmax-1)\suc+\adjsuc) > 1$. It then follows from Eq.~\eqref{eq: der efficm OR LOR at 1} that
\begin{equation}
\label{eq: der efficm OR LOR non-neg}
\frac{\partial \efficm}{\partial \RR} \geq 0 \quad \text{for any } \RR \in (0,1].
\end{equation}
Similarly, for $\prmax < (\suc-\adjsuc)/(2\suc)$ the numerator of Eq.~\eqref{eq: der efficm OR LOR} is a downward-opening parabola. In this case Descartes' rule of signs implies that it has one negative and one positive root, and again Eq.~\eqref{eq: der efficm OR LOR at 1} ensures that Eq.~\eqref{eq: der efficm OR LOR non-neg} holds. Lastly, for $\prmax = (\suc-\adjsuc)/(2\suc)$ the numerator of Eq.~\eqref{eq: der efficm OR LOR} is a decreasing straight line with positive $\efficm$-intercept, and Eq.~\eqref{eq: der efficm OR LOR non-neg} follows once more from Eq.~\eqref{eq: der efficm OR LOR at 1}. Thus Eq.~\eqref{eq: der efficm OR LOR non-neg} is satisfied in all cases. In consequence, using Eq.~\eqref{eq: efficm OR LOR 2},
\begin{equation}
\label{eq: inf efficm, RR small, OR LOR}
\inf_{\RR \in (0,1]} \efficm = \lim_{\RR \rightarrow 0} \efficm = \frac{1-\prmax}{\suc+\adjsuc}.
\end{equation}

For $\RR > 1$, instead of carrying out a similar analysis to obtain $\inf_{\RR \in (1,\infty)} \efficm$, it suffices to note that the right-hand side of Eq.~\eqref{eq: efficm OR LOR} is unchanged if $\RR$ is replaced by $1/\RR$ and $\adjsuc$ is replaced by $-\adjsuc$. Applying this transformation in Eq.~\eqref{eq: inf efficm, RR small, OR LOR} gives the result
\begin{equation}
\label{eq: inf efficm, RR large, OR LOR}
\inf_{\RR \in [1,\infty)} \efficm = \frac{1-\prmax}{\suc-\adjsuc}.
\end{equation}
From Eqs.~\eqref{eq: inf efficm, RR small, OR LOR} and \eqref{eq: inf efficm, RR large, OR LOR} it is concluded that $\inf_{\RR \in (0,\infty)} \efficm = (1-\prmax)/(\suc+\adjsuc)$.
Therefore Eq.~\eqref{eq: efficm get OR LOR} holds, which establishes Eq.~\eqref{eq: efic OR LOR, indep of RR}.

The right-hand side of Eq.~\eqref{eq: efic OR LOR, indep of RR} decreases with $\prmax$, and tends to $(\suc-\cerr)/(\suc+\adjsuc)$ as $\prmax \rightarrow 0$. This proves the first inequality in Eq.~\eqref{eq: efic OR LOR, indep of RR, lim}, and then the second follows from Eq.~\eqref{eq: suc tarvar}.
\qed


\begin{thebibliography}{18}
\providecommand{\natexlab}[1]{#1}
\providecommand{\url}[1]{{#1}}
\providecommand{\urlprefix}{URL }
\providecommand{\doi}[1]{\url{https://doi.org/#1}}
\providecommand{\eprint}[2][]{\url{#2}}
 \bibcommenthead

\bibitem[{Agresti(2002)}]{Agresti02}
Agresti A (2002) Categorical Data Analysis, 2nd edn. John Wiley and Sons

\bibitem[{Armitage et~al.(2002)Armitage, Berry, and Matthews}]{Armitage02}
Armitage P, Berry G, Matthews NS (2002) Statistical Methods in Medical
  Research, 4th edn. Blackwell

\bibitem[{Athreya and Lahiri(2006)}]{Athreya06}
Athreya KB, Lahiri SN (2006) Measure Theory and Probability Theory. Springer

\bibitem[{Cho(2019)}]{Cho19}
Cho H (2019) Two-stage procedure of fixed-width confidence intervals for the
  risk ratio. Methodology and Computing in Applied Probability 21(3):721--733.
  \doi{10.1007/s11009-019-09717-5}

\bibitem[{Cho and Wang(2020)}]{Cho20}
Cho H, Wang Z (2020) On fixed-width confidence limits for the risk ratio with
  sequential sampling. American Journal of Mathematical and Management Sciences
  39(2):166--181. \doi{10.1080/01966324.2019.1679301}

\bibitem[{Elias(1972)}]{Elias72}
Elias P (1972) The efficient construction of an unbiased random sequence.
  Annals of Mathematical Statistics 43(3):865--870.
  \doi{10.1214/aoms/1177692552}

\bibitem[{Haldane(1945)}]{Haldane45}
Haldane JBS (1945) On a method of estimating frequencies. Biometrika
  33(3):222--225. \doi{10.2307/2332299}

\bibitem[{Kay(1993)}]{Kay93}
Kay SM (1993) Fundamentals of Statistical Signal Processing: Estimation Theory,
  2nd edn. Prentice Hall

\bibitem[{Kokaew et~al.(2023)Kokaew, Bodhisuwan, Yangb, and Volodin}]{Kokaew23}
Kokaew A, Bodhisuwan W, Yangb SF, et~al (2023) Logarithmic confidence
  estimation of a ratio of binomial proportions for dependent populations.
  Journal of Applied Statistics 50(8):1750--1771.
  \doi{10.1080/02664763.2022.2041566}

\bibitem[{Komornik(2006)}]{Komornik06}
Komornik V (2006) Another short proof of {Descartes}'s rule of signs. The
  {American} Mathematical Monthly 113(9):829--830.
  \doi{10.1080/00029890.2006.11920371}

\bibitem[{Lehmann and Casella(1998)}]{Lehmann98}
Lehmann EL, Casella G (1998) Theory of Point Estimation, 2nd edn. Springer

\bibitem[{Mendo(2025)}]{Mendo25b}
Mendo L (2025) Estimating odds and log odds with guaranteed accuracy.
  Statistical Papers 66(1):1--17. \doi{10.1007/s00362-024-01639-w}

\bibitem[{Mendo(2026)}]{Mendo25c}
Mendo L (2026) Estimation of relative risk, odds ratio and their logarithms
  with guaranteed accuracy and controlled sample size ratio. Statistical Papers
  67(3):1--55. \doi{10.1007/s00362-026-01803-4}

\bibitem[{von Neumann(1951)}]{vonNeumann51}
von Neumann J (1951) Various techniques used in connection with random digits.
  National Bureau of Standards Applied Mathematics Series 12:36--38

\bibitem[{Paes~Leme and Schneider(2023)}]{PaesLeme23}
Paes~Leme R, Schneider J (2023) Multiparameter {Bernoulli} factories. Annals of
  Applied Probability 33(5):3987--4007. \doi{10.1214/22-AAP1913}

\bibitem[{Peres(1992)}]{Peres92}
Peres Y (1992) Iterating von {Neumann}'s procedure for extracting random bits.
  Annals of Statistics 20(1):590--597. \doi{10.1214/aos/1176348543}

\bibitem[{Pocock(1977)}]{Pocock77}
Pocock SJ (1977) Group sequential methods in the design and analysis of
  clinical trials. Biometrika 64(2):191--199. \doi{10.2307/2335684}

\bibitem[{Siegmund(1982)}]{Siegmund82}
Siegmund D (1982) A sequential confidence interval for the odds ratio.
  Probability and Mathematical Statistics 2(2):149--156

\end{thebibliography}

\end{document}